\documentclass[a4paper,11pt]{article}
\pdfoutput=1 

\usepackage{jcappub} 
\usepackage[title]{appendix}
\usepackage{subfigure}
\usepackage{orcidlink}
\usepackage{ulem} 
                     
\usepackage[T1]{fontenc} 

\def\md{\mathrm{d}}

\title{\boldmath Do JWST reionization (optical depth) puzzle, cosmological tensions, and CMB anomalies imply Harrison-Zel'dovich spectrum?}

\author[a,1]{Hao-Hao Li\note{Corresponding author.} \orcidlink{0000-0003-0974-771X},}
\author[a]{Xin-zhe Zhang \orcidlink{0000-0002-3264-7402},}
\author[a]{Taotao Qiu,}
\author[b,c]{Jun-Qing Xia}

\affiliation[a]{School of Physics, Huazhong University of Science and Technology,\\Luoyu Road 1037, Wuhan, China}
\affiliation[b]{School of Physics and Astronomy, Beijing Normal University, Beijing 100875, P.R.China}
\affiliation[c]{Institute for Frontiers in Astronomy and Astrophysics, Beijing Normal University, \\Beijing 100875, China}

\emailAdd{lihaohao23@hust.edu.cn}
\emailAdd{zincz@hust.edu.cn}
\emailAdd{qiutt@hust.edu.cn}
\emailAdd{xiajq@bnu.edu.cn}

\abstract{The James Webb Space Telescope (JWST) has observed massive galaxies at high redshifts, which implies an earlier epoch of reionization (EoR) compared with the cosmic microwave background (CMB) results. In this paper,  based on \texttt{Planck 2020} (NPIPE release), \texttt{ACT DR4} and \texttt{SPT-3G} data, if assumed a Harrison-Zel'dovich (HZ) primordial power spectrum in the standard cosmological model, we show that the redshift or optical depth of reionization is larger than the case of a power-law (PL) primordial power spectrum. In HZ-$ \Lambda $CDM model, the redshift of reionization is $ z_\text{reio} = 9.11 \pm 0.61 $, which is consistent with the JWST result that $ z_\text{reio} \approx 8.9 $. Moreover, the cosmological tensions, i.e. Hubble ($H_0$) tension and $ S_8 $ tension are alleviated in  HZ-$ \Lambda $CDM case. The Hubble constant is $ H_0 = 70.38 \pm 0.35 \, \text{km}/\text{s}/\text{Mpc}$ and the structure growth parameter is $ S_8 = 0.7645\pm 0.0094 $ in HZ-$ \Lambda $CDM model.  We also consider two extensions of $ \Lambda $CDM, including $ \Lambda $CDM$ + A_\text{L} $ and $ \Lambda $CDM$ + \Omega_\text{k} $ models. But the extensions of $ \Lambda $CDM with a HZ spectrum meet more serious CMB anomalies, i.e. lensing anomaly and spatial curvature anomaly as compared with the extensions of $ \Lambda $CDM with a PL spectrum. We discuss that these two CMB anomalies may come from the degeneracy of cosmological parameters.  }

\begin{document}
\maketitle
\flushbottom

\section{Introduction}
\label{sec:intro}

Recently, the JWST has observed bright galaxies at high redshifts, offering an indication of early star and galaxy formation than previously expected \cite{Donnan:2022fdr,Labbe:2022ahb,Boylan-Kolchin:2022kae}.  This observational result shows a reionization or optical depth puzzle compared with CMB results. The redshift  (or optical depth) of reionization in Planck 2018 result is  $ z_\text{reio} = 7.67 \pm 0.73 $ ( $ \tau_\text{reio} = 0.054 \pm 0.007 $)\cite{Planck:2018vyg}.  In the parametrized instantaneous reionization model, the reionization began at $  z < 10 $ and ended at $z \approx 6$. Since the formation of early stars and galaxies could ionize the intergalactic medium (IGM), this low-$z$ EoR is hard to explain the detection by JWST that there are some massive and bright galaxies at $ z > 10 $ \cite{Adams:2024wre,Finkelstein:2024wre,Donnan:2024mon}. 

 Before JWST, it is difficult to accurately describe the early stage of EoR. There are many elements that can affect the descriptions of reionization history in the traditional scenarios, for example, the ionizing efficiencies per unit UV luminosity $\xi_\text{ion}$ and the escape fraction $f_\text{esc}$ of ionized photons \cite{Munoz:2024fas, Melia:2024buj, Robertson:2021ljt,Qin:2020xrg,Cain:2024fbi}. Some recent works show that JWST measurements imply a higher value of the ionizing efficiencies $\xi_\text{ion}$ and an earlier evolution of the galaxy UV luminosity function (LF) \cite{Adams:2024wre,Donnan:2024mon,Simmonds:2024nov}. These results combined with some observationally resonable assumptions about low-$z$  ionizing escape fractions suggest the reionization ended at $z \sim 9 $, inconsistent with  Planck 2018 measurement nearly $ 2 \sigma $ level \cite{Munoz:2024fas}. A number of possible resolutions have been proposed to prove an early EoR, such as assuming high star formation efficiency, constructing new  descriptions of the UV luminosity and so on \cite{Pallottini:2023yqg, Dekel:2023ddd, Harikane:2022rqt, Ferrara:2022dqw, Iocco:2024rez, Ziparo:2022rir}.

This high-$z$ EoR has a significant impact on the detection of cosmological parameters. There should be a higher redshift of reionization than previous CMB results, in other words, a larger optical depth is needed in cosmological analyses \cite{Li:2024rgq, Giare:2023ejv, Sokoliuk:2025phe,Forconi:2023izg}. In this work, we find that the cosmological model $ \Lambda $-Cold Dark Matter ($ \Lambda $CDM)  with a Harrison-Zel'dovich primordial power spectrum (HZ-$ \Lambda $CDM) as initial conditions will give a larger optical depth, which is compatible with the JWST observation.  The HZ spectrum  means that the spectral index $ n_\text{s} = 1 $, also named the scale-invariant spectrum, which could be associated with a exactly de-Sitter primordial inflation epoch. There is about $ 8 \sigma$ away from the scale-invariance in Planck 2018 result\cite{Planck:2018vyg}, which confirms the red tilt of the spectrum  as $ n_\text{s} = 0.9649 \pm 0.0042 $, but some observations give priority to a spectral index of unity \cite{ACT:2020gnv, Montero-Camacho:2024xvf, Giare:2022rvg}.  This does not rule out the possibility of the scale-invariant spectrum.

The primordial scalar power spectrum with the power-law (PL) form can be parameterized as 
\begin{equation}
	P_\text{s} = A_\text{s}\left( \frac{k}{k_\ast} \right)^{n_\text{s}-1}, \label{PL-PS}
\end{equation}
where $A_\text{s}$ is the amplitude of the primordial scalar power spectrum, $n_\text{s}$ is the spectral index, and $ k_\ast = 0.05 \, \text{Mpc}^{-1} $ is the pivot scale in this condition. Notice that the spectral index is unity in HZ spectrum.

Ground-based CMB observations prefer a HZ spectrum, especially the high-$\ell$ data \cite{ Giare:2022rvg, Hazra:2024nav, SPT-3G:2022hvq}. The DR4 data from Atacama Cosmology Telescope (ACT) give that $ n_\text{s} = 1.008 \pm 0.015 $, but when combined with the space-based CMB observations which include low-$\ell$ anisotropies and polarizations, such as WMAP or Planck, the spectral index become $n_\text{s} = 0.9729 \pm 0.0061$ (ACT$+$WMAP) and $n_\text{s} = 0.9691 \pm 0.0041$ (ACT$+$Planck) \cite{ACT:2020gnv}. The value of spectral index from SPT-3G 2018 is $ n_\text{s} = 0.970 \pm 0.016 $, but when excising the part of $\ell < 1000$ of the temperature power spectrum, it gives $ n_\text{s} = 0.994 \pm 0.018 $ \cite{SPT-3G:2022hvq}. In addition, lack of the large-scale CMB polarizations tends to give a higher value of the optical depth which could significantly alleviate or even eliminate CMB anomalies, since optical depth consistently exhibits correlations with anomalous parameters, such as $A_\text{L}$ and $\Omega_\text{k}$ \cite{Giare:2023ejv}. These discoveries imply that the spectral index, optical depth,  lensing amplitude, and curvature density are correlated with each other, especially reflected by low-$\ell$ anisotropies and polarizations.  We will provide a discussion about the degeneracies of these parameters. 

On the other hand, the standard cosmological model $ \Lambda $CDM with a PL primordial power spectrum (PL-$ \Lambda $CDM) faces many challenges like Hubble ($H_0$) tension,  $ S_8 $ tension and others \cite{Perivolaropoulos:2021jda}. A lot of solutions to these challenges tend to the HZ spectrum,  such as the primordial bounce-inflation scenario, the early dark energy scenario \cite{Li:2024rgq, Jiang:2022qlj, Wang:2024tjd, Jiang:2023bsz}. In the HZ-$ \Lambda $CDM model, cosmological tensions, i.e. $H_0$ tension and $ S_8 $ tension, are alleviated. Assuming a flat universe, CMB data give that $ H_0 = 70.38 \pm 0.35 \, \text{km}/\text{s}/\text{Mpc}$, which reduces the Hubble tension to about $ 2.5 \sigma $ compared with local observations that $ H_0 = 73.04 \pm 1.04 \, \text{km}/\text{s}/\text{Mpc}$  \cite{Riess:2021jrx} and has no tension with the result from Chicago-Carnegie Hubble Program (CCHP) that $ H_0 = 69.96 \pm 1.05 \, \text{km}/\text{s}/\text{Mpc}$ \cite{Freedman:2024eph}.  In addition, the structure growth parameter is $ S_8 = 0.7645\pm 0.0094 $ in HZ-$ \Lambda $CDM by using CMB data, which is consistent with the late universe observations \cite{KiDS:2020suj,Heymans:2020gsg}.
 
CMB lensing anomaly and the spatial curvature anomaly are two of the most serious CMB anomalies. CMB gravitational lensing occurs as CMB photons are deflected by the large-scale distribution of matter traveling through the universe. This effect leaves subtle imprints in the temperature and polarization anisotropies of CMB, which can be used to reconstruct a map of the lensing potential. The power spectrum of the lensing potential can be rescaled by a phenomenological parameter, the gravitational lensing amplitude $ A_\text{L} $. In $\Lambda$CDM model, $ A_\text{L}$  is predicted to be unity. However, the results of Planck satellite shows $ A_\text{L} = 1.18 \pm 0.065 $, which deviates from unity at nearly $ 3 \sigma $ level \cite{Planck:2018vyg, Domenech:2019cyh}. This result implies that the lensing effect is stronger than the prediction of $\Lambda$CDM model. This deviation is named CMB lensing anomaly.In this work,  we find that $ A_\text{L} = 0.997 \pm 0.040 $ with CMB data in the extended  PL-$\Lambda$CDM$ + A_\text{L}$ model and $ A_\text{L} = 1.124 \pm 0.042 $ with CMB data in the HZ-$\Lambda$CDM$ + A_\text{L}$ model. The lensing amplitude with a PL primordial spectrum is closer to unity, which gives negative support to HZ-$ \Lambda $CDM.  However, we speculate that this negative case is due to the degeneracy among the parameters and give a discussion in Section \ref{Sec:discussion}.

 As for the spatial curvature density anomaly, observations from the Planck satellite imply a negative value of  the spatial curvature density parameter, that $ \Omega_\text{k} =-0.0106\pm0.0065 $ (68\%, \texttt{Planck2018 TT,TE,EE+lowE+lesing}). This means that the universe is closed \cite{Planck:2018vyg,DiValentino:2019qzk} rather than flat predicted by PL-$ \Lambda $CDM. However, many local observations show evidence of a flat universe \cite{DESI:2024mwx,Bargiacchi:2021hdp,Efstathiou:2020wem,Vagnozzi:2020rcz,Vagnozzi:2020dfn,Dhawan:2021mel,Jiang:2024xnu}.  In this work, $ \Omega_\text{k} = -0.0011^{+0.0074}_{-0.0061} $ from CMB data in the extended  PL-$\Lambda$CDM$ + \Omega_\text{k}$ model and $ \Omega_\text{k} = -0.0204^{+0.0091}_{-0.0075} $ with CMB data in  HZ-$\Lambda$CDM$ + \Omega_\text{k}$ model.  Adding BAO measurements with CMB data to break the geometrical degeneracy, we have $ \Omega_\text{k} = 0.0019\pm 0.0014 $ in  PL-$\Lambda$CDM$ + \Omega_\text{k}$ model and $ \Omega_\text{k} = -0.0035\pm 0.0012 $ in  HZ-$\Lambda$CDM$ + \Omega_\text{k}$ model. If our universe is flat, similar to the lensing anomaly, our results also give negative support to HZ-$ \Lambda $CDM.

 Even though, HZ-$\Lambda$CDM gets supports by Hubble constant observations and the reionization optical depth result from JWST. Due to the suppression on the large scale angular power spectrum of CMB anisotropies and the enhancement on the small scale from a larger spectral index $n_{\text{s}}$ showed in Figure \ref{fig:DlTT_Delta}, $\tau_{\text{reio}}$ in CMB estimations should increase to push down the small scale enhancement, so that a larger $z_{\text{reio}}$. Thus, the larger observed $z_{\text{reio}}$ could indicate some physics like HZ spectrum from a de Sitter primordial universe.

We use the package \textsf{CAMB}\footnote{\url{https://github.com/cmbant/camb}} as the Einstein-Boltzmann equation solver and the package \textsf{Cobaya}\footnote{\url{https://github.com/CobayaSampler/cobaya}} \cite{Torrado:2020dgo} as the Markov Chain Monte Carlo(MCMC) sampler to product the posterior estimate of cosmological parameters. The Gelman-Rubin criterion for all chains is converged to $ R-1 < 0.01 $. To analyze the chains and plot the posterior distributions, we use \textsf{GetDist}\footnote{\url{https://github.com/cmbant/getdist}} \cite{Lewis:2019xzd}.

 The data sets we used are the following. 
 \\
\textbf{CMB data} -- \texttt{Planck 2018 lowl\_TT} \cite{Planck:2018vyg}, \texttt{Planck 2020 lollipop lowl\_E} and \texttt{hillipop TTTEEE} (NPIPE release) \cite{Planck:2020olo,Tristram:2023haj}. Hereafter, all the Planck data are labeled as \texttt{Planck 2020}.  Other CMB data we used include \texttt{the Data Release 4 by the Atacama Cosmology Telescope} (hereafter, \texttt{ACT DR4}) \cite{ACT:2020gnv} 
and \texttt{SPT-3G 2018 TTTEEE} (hereafter, \texttt{SPT3G}) \cite{SPT-3G:2022hvq}. There is a subset of the \texttt{ACT DR4} likelihood -- the \texttt{pyactlike.ACTPol\_lite\_DR4\_for\_combining\_with\_Planck} for TT,TE,EE excluding the large scale ACT temperature when combining with Planck data. However,  we use the total likelihood \texttt{pyactlike.ACTPol\_lite\_DR4} here. There are some discrepancies between the Planck, ACT and SPT data sets, as shown in \cite{Calderon:2023obf}, but we don't apply cuts on the multipole ranges of different CMB data sets to avoid overlaps for keeping generalities.
\\
\textbf{BAO data} -- Baryon acoustic oscillations (BAO) measurement from the first year of observations from \texttt{the Dark Energy Spectroscopic Instrument} (DESI), (hereafter, \texttt{DESI BAO}) \cite{DESI:2024mwx}. 

\section{JWST reionization (optical depth) puzzle}
\label{Sec:JWST-puzzle}

The optical depth to reionization is defined by
\begin{equation}
	\tau_{\text{reio}} (z_{\ast}) = \int_{0}^{z_{\ast}} n_\text{e} (z) \sigma_\text{T} \frac{c}{(1+z)H(z)} \md z, \label{def-tau}
\end{equation}
where $n_\text{e}(z) $ is the number density of free electrons produced by reionization and $\sigma_\text{T} = 6.65 \times 10^{-29}$m$^2$ is the Thomson scattering cross-section. After recombination, the ionization fraction $ X_\text{e} \simeq 0$ until cosmic dawn at $ z \sim 30  $ to $ 50 $, hence the upper limit of integral can be set as $ z_\ast = 50 $.

 The above integral can be done approximately by adopting a standard $ \Lambda $CDM cosmology. In a flat universe and ignoring the radiation, the first Friedmann equation is $H(z)=H_0\left[\sqrt{\Omega_\text{m}(1+z_{\text{reio}})^3+\Omega_\Lambda-1} \right]$ with $\Omega_\text{m}+\Omega_\Lambda=1$. If we define the redshift of reionization by half-ionized hydrogen and helium, we get \cite{Shull:2008su}
\begin{equation}
	\tau (z_\text{reio}) \approx 0.039 \frac{\Omega_\text{b} h}{\Omega_\text{m}} \left( \sqrt{\Omega_\text{m} (1+z_\text{reio})^3 + \Omega_\Lambda} -1 \right) . \label{def-tau-approx}
\end{equation}
For large redshifts $\left( z > 2 \right)$ , $ \Omega_\text{m} (1+z)^3 \gg \Omega_\Lambda $, the above equation is simplified to
\begin{equation}
	\tau_\text{reio} \sim 2.3 \times 10^{-3} \left(1+z_\text{reio} \right)^{3/2},  \label{tau-approx}
\end{equation}
with some estimates, $ h \approx 0.7 $, $ \Omega_\text{m} \approx 0.28 $, and $ \Omega_\text{b} \approx 0.045 $. We can estimate the redshift of reionization $ z_\text{reio} $ by using equation \eqref{tau-approx} at large redshift.

The ionization fraction of reionization is $ X_\text{e}(z)  = n_\text{e}(z)/n_\text{H}(z) $ and $n_\text{H}(z) = \left\langle n_\text{H} \right\rangle \left(1+z \right)^{3} $ is the number density of hydrogen with the mean comoving number density of hydrogen $\left\langle n_\text{H} \right\rangle = Y_\text{H} \Omega_\text{b} \rho_\text{c} / m_\text{H} $, which depends on the primordial mass fraction of hydrogen $ Y_\text{H} $,  the baryon density parameter $ \Omega_\text{b} $,  the critical density $ \rho_\text{c} $ and the mass of hydrogen $ m_\text{H} $.  In CMB numerical analyses, the EoR is usually parameterized by a \textit{Tanh} model \cite{Lewis:2008wr} like
\begin{equation}
	X_\text{e}(z)  = \frac{f}{2}\left\lbrace 1 + \tanh \left[\left(  1-\left( \frac{1+z}{1+z_\text{reio}}\right)^{3/2}\right) \left( \frac{1+z_\text{reio}}{1.5\Delta z}\right)  \right]  \right\rbrace, 
\end{equation}
with
\begin{equation}
	f = \begin{cases}
		1 + X_\text{He}, \,\, z > 3.5 \\ 
		1 + 2X_\text{He},\,\, z < 3.5
	\end{cases},
\end{equation}
where $ X_\text{He} = n_\text{He}/n_\text{H} $ is the ionized fraction contributed by helium and $ n_\text{He} $ is the number density of helium. Notice that helium is assumed to be singly ionized with hydrogen at $ z > 3.5 $. Approximately, the mass fraction of helium is $ Y_\text{He} =1 - Y_\text{H} \approx 0.24 $, therefore $ f \approx 1 + 0.24/4/(1-0.24) = 1.08 $ with singly ionized helium at  higher redshift and $ f \approx 1+ 0.24/2/(1-0.24) = 1.16 $ with doubly ionized helium at lower redshift \cite{Lewis:2008wr}. Usually, the instantaneous transition from singly ionized helium to doubly ionized helium is also assumed as a \textit{Tanh} model, although it adds a tiny contribution to low-redshift reionization.

\begin{figure}[tbp]
	\centering
	\includegraphics[width=1.0\textwidth]{{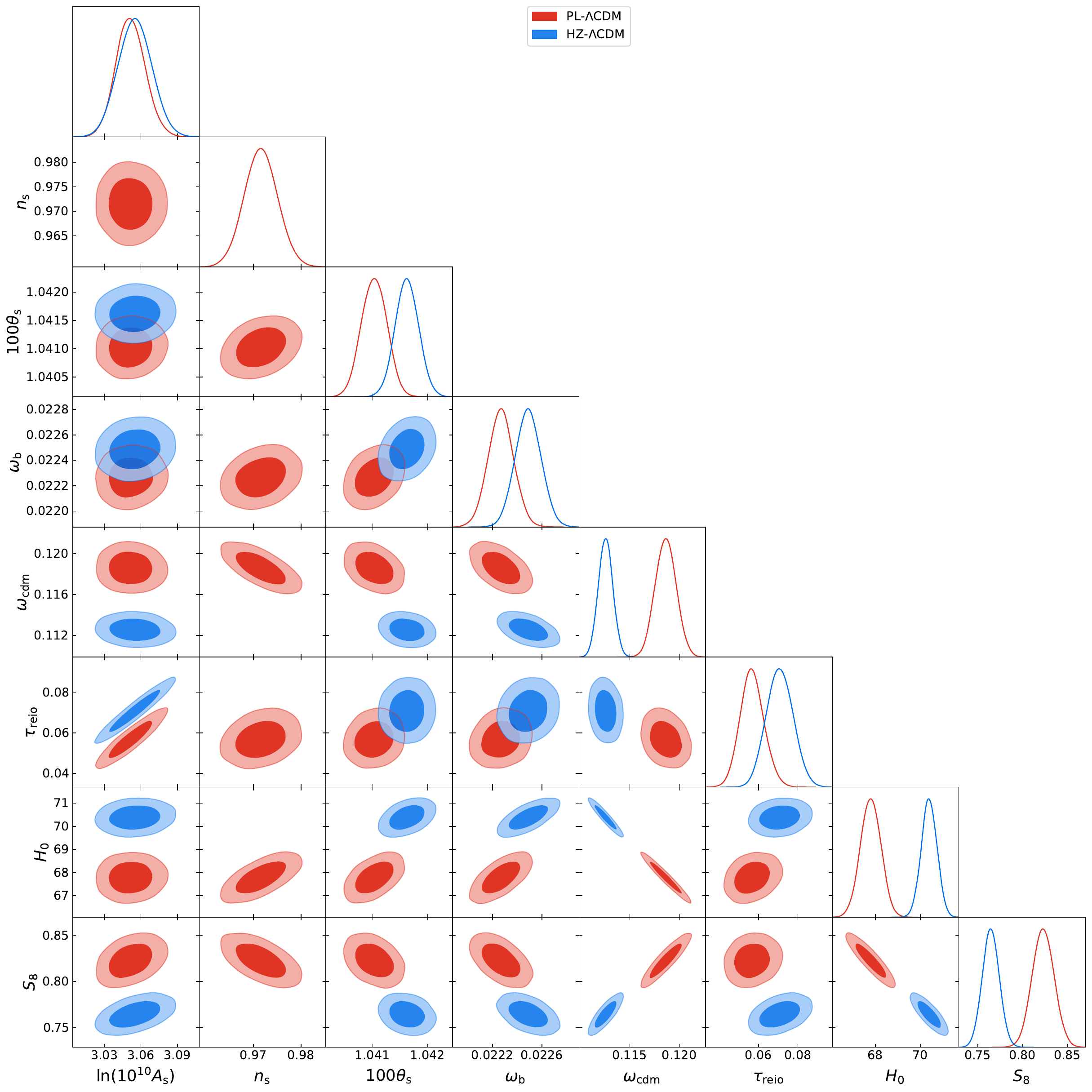}}
	\caption{\label{fig:LCDM_All} Triangle plot for cosmological parameters with CMB data in PL/HZ-$\Lambda$CDM model.  }
\end{figure}

Cosmological parameters yielded by CMB observations are shown in Table \ref{tab-LCDM}. The redshift (or optical depth) of reionization from CMB data is  $ z_\text{reio} = 7.93 \pm 0.60 $ ( $ \tau_\text{reio} = 0.0568^{+0.0056}_{-0.0063} $) in the PL-$ \Lambda $CDM model, which is almost no difference from the results of Planck 2018 \cite{Planck:2018vyg}. But in the HZ-$ \Lambda $CDM, there is an earlier EoR. The redshift (or optical depth) of reionization from CMB data is  $ z_\text{reio} = 9.11 \pm 0.61 $ ( $ \tau_\text{reio} = 0.0709\pm 0.0067 $). The contour plots of cosmological parameters are shown in Figure \ref{fig:LCDM_All}.  A large optical depth can wash out the primordial anisotropies. However, a change in the amplitude of primordial spectrum $ A_\text{s} $, together with the spectral index $ n_\text{s} $, can mimic the effect of optical depth, thus there is strong degeneracy between \{$ A_\text{s} $, $ n_\text{s} $\} and $ \tau_\text{reio} $ \cite{Dodelson:2020bqr}. We will discuss this degeneracy more concretely in Section \ref{Sec:discussion}.

\begin{table}[htbp]
	\centering
	\begin{tabular} {| l | c | c | }
		\hline
		Parameter &  PL-$\Lambda$CDM   &  HZ-$\Lambda$CDM \\
		\hline
		{$\ln(10^{10} A_\text{s})$} & $3.052\pm 0.012            $ & $3.055\pm 0.014            $\\
		
		{$n_\text{s}     $} & $0.9716\pm 0.0035          $ &                             \\
		
		{$100\theta_\text{s}$} & $1.04103\pm 0.00023        $ & $1.04162\pm 0.00021        $\\
		
		{$\omega_\text{b}$} & $0.02227\pm 0.00010        $ & $0.02249\pm 0.00010        $\\
		
		{$\omega_\text{cdm}$} & $0.1186\pm 0.0010          $ & $0.11257\pm 0.00072        $\\
		
		{$\tau_\text{reio}$} & $0.0568^{+0.0056}_{-0.0063}$ & $0.0709\pm 0.0067          $\\
		\hline
		$H_\text{0}                $ & $67.79\pm 0.45             $ & $70.38\pm 0.35             $\\
		
		$\Omega_\text{m}           $ & $0.3080\pm 0.0062          $ & $0.2740\pm 0.0040          $\\
		
		$z_\text{reio}             $ & $7.93\pm 0.60              $ & $9.11\pm 0.61              $\\
		
		$S_\text{8}                $ & $0.823\pm 0.012            $ & $0.7645\pm 0.0094          $\\
		\hline
	\end{tabular}
	\caption{\label{tab-LCDM} Cosmological parameters with mean value and $ 68\% $ limits by using CMB data (\texttt{Planck 2020} + \texttt{ACT DR4} + \texttt{SPT3G}  ) in PL/HZ-$\Lambda$CDM model.  A larger optical depth or redshift of reionization in the HZ-$\Lambda$CDM model shows an earlier EoR.}
\end{table}

Another way to describe the EoR is by calculating the ionized fraction of hydrogen from local observations. The ionized fraction of hydrogen is defined as $ X_\text{HII} \equiv n_\text{HII}/n_\text{H} = X_\text{e}/f $. This quantity evolves as \cite{Robertson:2013bq}
\begin{equation}
	\dot{X}_\text{HII} = \frac{\dot{n}_\text{ion}}{\left\langle n_\text{H}\right\rangle} - \frac{X_\text{HII}}{t_\text{rec}} .
\end{equation}
The IGM recombination time can be written as
\begin{equation}
	t_\text{rec} = \left[\left(1+ X_\text{He} \right)* C  \alpha_\text{B} n_\text{H} \right]^{-1}, 
\end{equation}
where $ \alpha_\text{B} $ is the case B recombination coefficient at an IGM temperature $ T = 20000 $K and the clumping factor that accounts for the effects of IGM inhomogeneity can be set $  C = 3 $ followed by previous works \cite{Pawlik:2008mr,Shull:2011aa}. 

If the ionizing photons are from star-forming galaxies, the time-dependent cosmic ionization rate is \cite{Munoz:2024fas,Melia:2024buj}  
\begin{equation}
	\dot{n}_\text{ion} = \int_{-\infty}^{M_\text{UV,cutoff}} \md M_\text{UV}\Phi_\text{UV} \dot{N}_\text{ion} f_\text{esc}.
\end{equation}
The cutoff magnitude used here is the same as \cite{Melia:2024buj}, $M_\text{UV,cutoff} = -14.6$ for $ \Lambda $CDM. $\Phi_\text{UV}$ is the UV luminosity function (UVLF), often assumed to be a Schechter form \cite{Schechter:1976iz}
\begin{equation}
	\Phi_\text{UV}\left( M_\text{UV} \right) = \left( 0.4\ln 10 \right) \phi_\star\left[10^{0.4\left( M_\star -M_\text{UV}\right)}  \right]^{1+\alpha} \exp\left[-10^{0.4\left( M_\star -M_\text{UV}\right)}  \right].  
\end{equation}
The UVLF describes the number of galaxies per unit volume as a function of their UV luminosity, typically defined at a rest-frame wavelength of 1500 \AA. The UV light emitted from galaxies is primarily produced by young, hot stars, and thus the UVLF provides insights into the star formation rates and the properties of stellar populations within galaxies. It depends on three parameters, the normalization parameter $ \phi_\star $ $\left( \text{in units of Mpc}^{-3} \text{mag}^{-1} \right) $, the characteristic galaxy magnitude $ M_\star $, and the faint-end slope $ \alpha $. We take this UVLF from the pre-JWST fit in \cite{Bouwens:2021AJ} at $ z < 9 $ and from the JWST calibrations of \cite{Donnan:2024mon} at $ z > 9 $. 

The $ f_\text{esc} $ is the fraction of photons produced by stellar populations that escape to ionize the IGM. It is impossible to constrain this escaping fraction from observations at high redshift because the escaping ionizing photons are absorbed by IGM before reaching the earth. However, a low redshift (low-$z$) analogues was fitted from \cite{Chisholm:2022mnras}
\begin{equation}
	f_\text{esc} = A_\text{f} \times 10^{b_\text{f} \beta_\text{UV}},
\end{equation}
with $ A_\text{f} = 1.3 \times 10^{-4} $ and $ 10^{b_\text{f}} = -1.22 $. $ \beta_\text{UV} $ is the UV slope. To avoid extrapolating $ f_\text{esc} $ in this relation, we can take $ \beta_\text{UV} = -2.7 $\cite{Munoz:2024fas,Melia:2024buj}.

$ \dot{N}_\text{ion} \equiv L_\text{UV} \xi_\text{ion} $ is the ionizing-photon production rate per galaxy, in terms of the UV luminosity $ L_\text{UV} $. $ \xi_\text{ion} $ is the ionizing efficiency which counts the ionizing photons per unit UV luminosity. JWST  opens a new window to detect reionization. The new data of JWST suggests a higher ionizing efficiency with high-redshift and brighter star-formation galaxies, which  is well fitted by \cite{Simmonds:2024nov}
\begin{equation}
	\log_\text{10} \left[ \xi_\text{ion}/\left(\text{Hz erg}^{-1} \right)  \right] \approx 0.05\left(z - 9 \right) + 0.11 \left(M_\text{UV} +16.5 \right) + 25.97. \label{eq:xi_ion}
\end{equation}
If we set $ z = 9 $ and  $M_\text{UV} =-16.5 $, then $ \log_\text{10} \xi_\text{ion} = 25.97 \left(  \text{Hz erg}^{-1}\right)  $. This value is greater than the canonical one previously used, that $ \log_\text{10} \xi_\text{ion} = 25.2 \left(  \text{Hz erg}^{-1}\right)  $ \cite{Robertson:2013bq, Bouwens:2016aa}. 

\begin{figure}[htbp]
	\centering 
	\subfigure{ \includegraphics[width=.95\textwidth]{{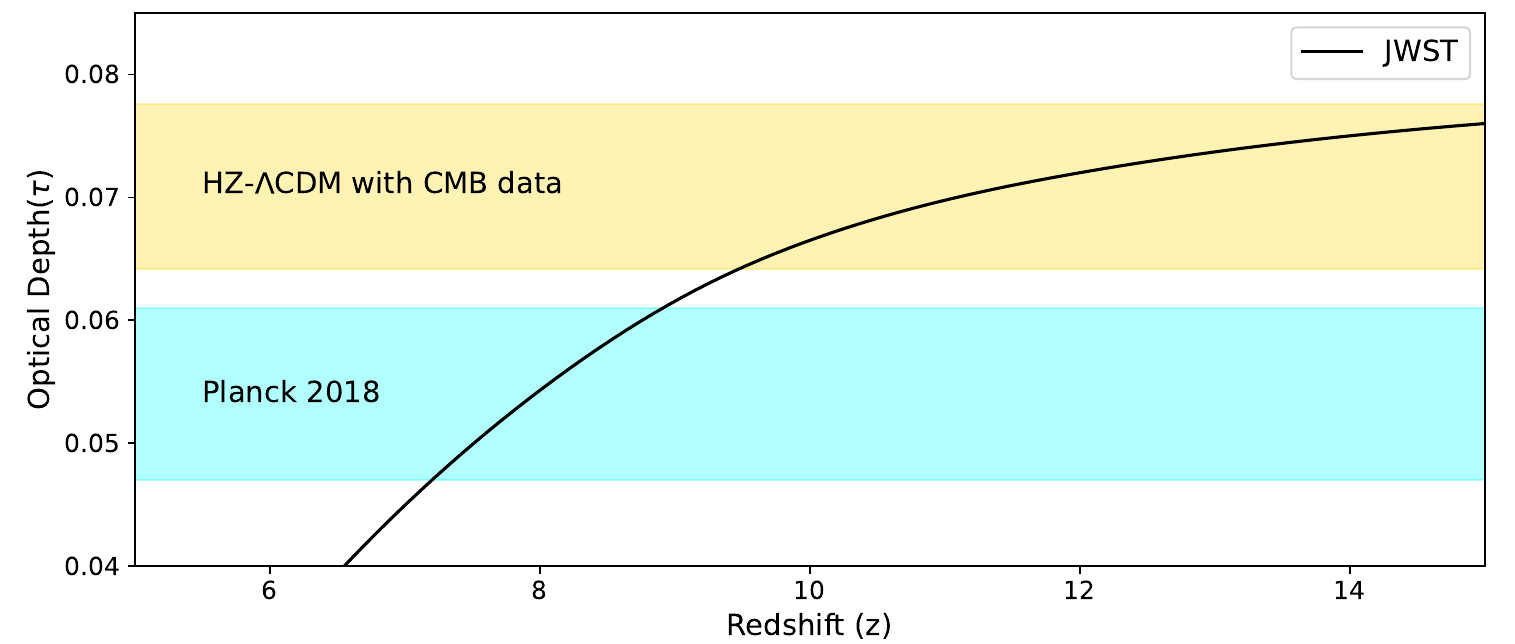}} }
	\subfigure{ \includegraphics[width=.95\textwidth]{{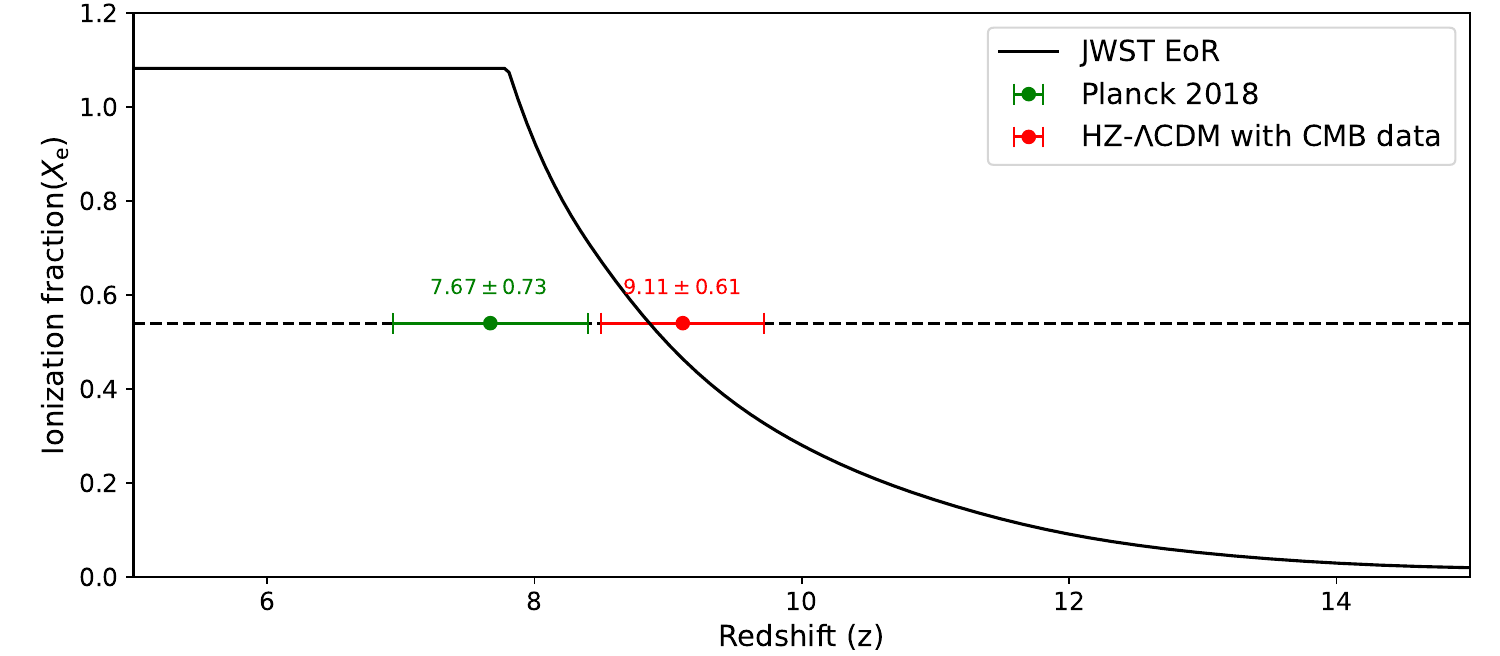}} }
	\caption{\label{fig:JWST} \textbf{Top:} The optical depth $ \tau $ as a function of redshift $z$.  The cyan color band  presents the mean value and $ 1\sigma $ limit of \texttt{Planck 2018 TTTEEE}, \texttt{lowT}, \texttt{lowE} and \texttt{lensing data}, which is $ \tau_\text{reio} = 0.054 \pm 0.007 $ \cite{Planck:2018vyg}. The gold color band is the mean value and $ 1\sigma $ limit of HZ-$\Lambda$CDM model with CMB data, which is $ \tau_\text{reio} = 0.0709 \pm 0.0067 $ in our work. The dark solid line based on new JWST observations, for a cutoff $ M_\text{UV,cutoff} = -14.6 $ and the JWST-calibrated ionizing efficiency, $ \xi_\text{ion} $, see equation \eqref{eq:xi_ion}.  \textbf{Bottom:} The ionization fraction $ X_\text{e}(z)$ from reionization described by JWST. The dashed line is the half of the largest value of $ X_\text{e} $ only considering totally ionized hydrogen and the singly ionized helium, that is $f/2 \approx 0.54$. The $z$ value of the intersection between the dashed line and JWST EoR is $ z_\text{reio,JWST} \approx 8.9 $, which is consistent with  $ z_\text{reio} = 9.11 \pm 0.61 $ at $1 \sigma$ level from HZ-$\Lambda $CDM constrained by CMB data. The green point with error bar is $ z_\text{reio} = 7.67 \pm 0.73 $ constrained by \texttt{Planck 2018 TT,TE,EE} and \texttt{lensing data} \cite{Planck:2018vyg}.  } 
\end{figure}

Following the upper definitions, equation \eqref{def-tau} can be wrote as
\begin{equation}
	\tau_{\text{reio}} (z) =c \sigma_\text{T}\left\langle n_\text{H}\right\rangle \int_{0}^{z} f X_\text{HII} (\tilde{z}) \frac{(1+\tilde{z})^2}{H(\tilde{z})} \md \tilde{z}. \label{astro-tau}
\end{equation}
The EoR can be described by the evolution of the ionized fraction of hydrogen $ \left(X_\text{HII} \right)  $ or the neutral fraction of hydrogen $ \left(X_\text{HI} = 1- X_\text{HII} \right)  $. 

We plot the optical depth as a function of redshift with constraints of new JWST observations of structure formation in the early universe, as shown in Figure \ref{fig:JWST}. In the top one, the optical depth described by JWST observations is  $\tau_{\text{reio,JWST}} \approx 0.075 $, which shows a $3 \sigma$ tension with the result $\tau_{\text{reio}}=0.054\pm0.007$ of Planck 2018 based on the PL-$\Lambda$CDM model \cite{Planck:2018vyg}, but is consistent with the result $\tau_{\text{reio}}=0.0709\pm0.0067$ of HZ-$\Lambda$CDM model constrained by CMB data  in this work at $0.61 \sigma$ level. In the bottom one, JWST observations show an earlier  midpoint of EoR with $z_{\text{reio,JWST}}\approx8.9$, which shows a $1.7 \sigma$ tension with the result $z_{\text{reio}}=7.67\pm0.73$ of Planck 2018 based on the PL-$\Lambda$CDM model \cite{Planck:2018vyg}, but is compatible with the result $z_{\text{reio}}=9.11\pm0.61$ of HZ-$\Lambda$CDM model in this work at $0.34 \sigma$ level.

The difference at $ 2 \sim 3 \sigma $ level in redshift of reionization or optical depth between JWST and CMB measurements in PL-$\Lambda$CDM is named JWST reionization (optical depth) puzzle. Our results in Table \ref{tab-LCDM} and Figure \ref{fig:JWST} show that this puzzle can be averted to less than $1 \sigma $ level by assuming a HZ primordial power spectrum in $\Lambda$CDM universe.  Using the recent CMB data to fit the cosmological parameters in PL-$\Lambda$CDM model, the results $\tau_\text{reio} = 0.0568^{+0.0056}_{-0.0063}$ ($z_\text{reio} = 7.93\pm 0.60 $) still show some tensions at $ 3.3 \sigma $ ($ 1.6 \sigma $) level compared with JWST results.

\section{Alleviating cosmological tensions with Harrison-Zel'dovich spectrum }
\label{Sec:tension}
\subsection{Hubble ($H_0$) tension}
\label{Sec:H0}

\begin{figure}[htbp]
	\begin{flushleft}
		\includegraphics[width=0.95\textwidth]{{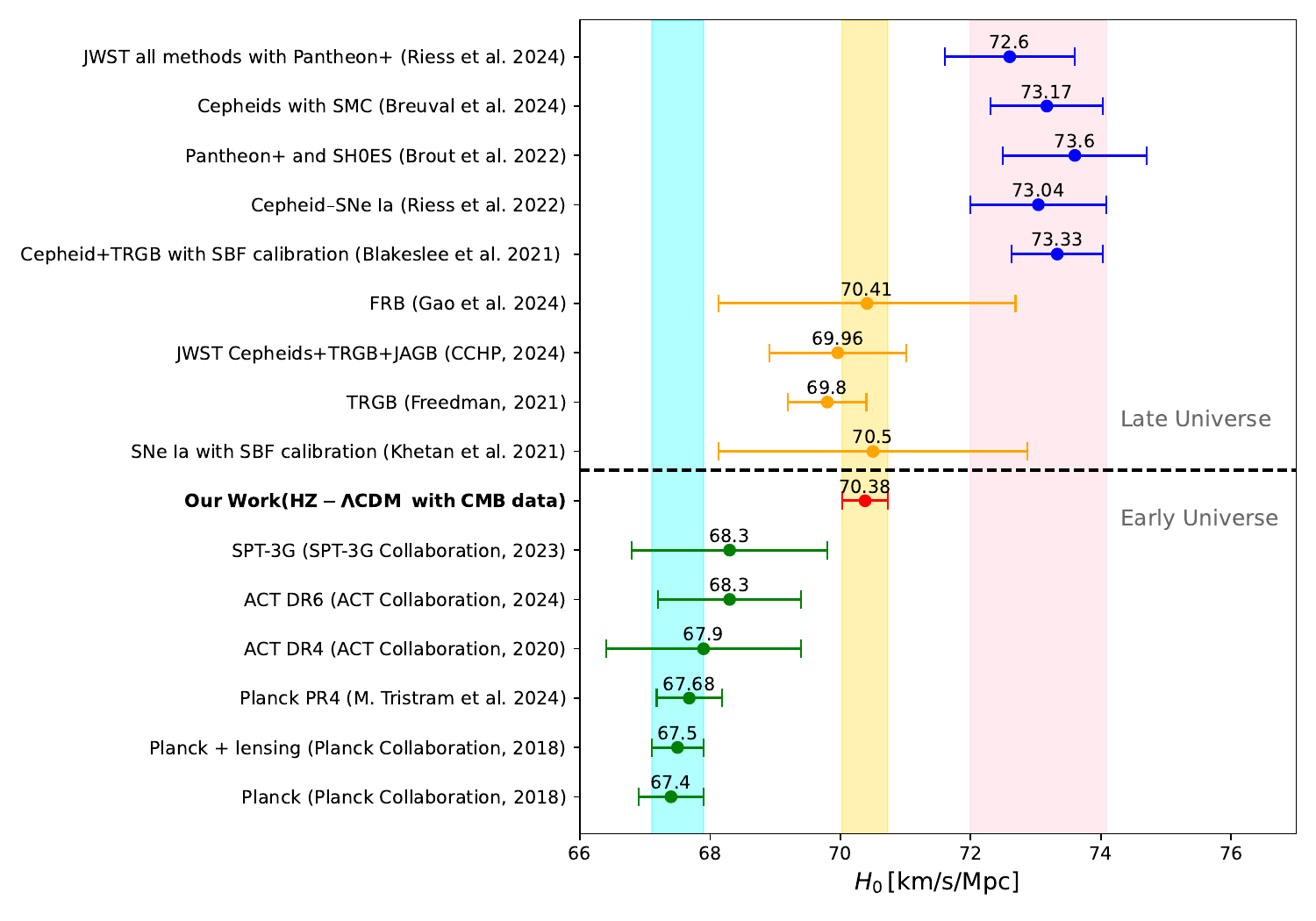}}
		\caption{\label{fig:H0} $H_0$ values of early and late observations. The cyan color band  presents the mean value and $ 1\sigma $ limit of \texttt{Planck 2018 TTTEEE}, \texttt{lowT}, \texttt{lowE} and \texttt{lensing data}, which is $ H_0 = 67.5 \pm 0.5 $ km/s/Mpc\cite{Planck:2018vyg}. The gold color band is the mean value and $ 1\sigma $ limit of HZ-$\Lambda$CDM model with CMB data, which is $ H_0 = 70.38 \pm 0.35 $ km/s/Mpc in our work. The pink color band presents the mean value and $ 1\sigma $ limit of Cepheid–SNe Ia sample, which is $ H_0 = 73.04 \pm 1.04 $ km/s/Mpc\cite{Riess:2021jrx}.  }
	\end{flushleft}
\end{figure}

Hubble tension is the most serious crisis in the standard cosmological model.  \texttt{Planck 2018} as one of the CMB observations shows that $ H_0 = 67.4 \pm 0.5\, \text{km}/\text{s}/\text{Mpc} $ \cite{Planck:2018vyg}, the local measurement of distance ladders constructed from Type Ia supernova (SNe Ia) gives $ H_0 = 73.04 \pm 1.04\, \text{km}/\text{s}/\text{Mpc}  $ \cite{Riess:2021jrx}. There is about $ 5\sigma $ tension of the measurements about $ H_0 $ between CMB under $\Lambda$CDM and local distance ladders \cite{SPT-3G:2022hvq,ACT:2020gnv,ACT:2023kun, Tristram:2023haj, Riess:2024vfa,Breuval:2024lsv, Brout:2022vxf, Blakeslee:2021rqi}, as shown in Figure \ref{fig:H0}. Recently, the crisis even increased to about $ 5.8\sigma $  by combining four geometric distances with the Hubble Space Telescope (HST) photometry of Small Magellanic Cloud (SMC) Cepheids \cite{Breuval:2024lsv}.

 In recent years, some local observations give a middle value of Hubble constant, $ H_0 \sim 70 \text{km}/\text{s}/\text{Mpc}  $ \cite{Gao:2024kkx,Freedman:2024eph, Freedman:2021ahq,Khetan:2020hmh}, especially the CCHP's result from JWST Cepheids + TRGB + JAGB that $ H_0 = 69.96 \pm 1.05 \, \text{km}/\text{s}/\text{Mpc}$ \cite{Freedman:2024eph}. In our numerical analysis, the Hubble constant $ H_0 = 70.38 \pm 0.35 \, \text{km}/\text{s}/\text{Mpc}$ in HZ-$\Lambda$CDM with CMB data. On the one hand, this result is still in tension with Cepheid-based measurements, but this tension is reduced to $2\sigma$ level. On the other hand, this result has no tension compared with CCHP's result, meanwhile the Hubble tension is still at $2\sigma$ level  between CMB data in PL-$\Lambda$CDM and measurements of CCHP group, see Table \ref{tab-LCDM} and Figure \ref{fig:H0}. 

Usually, the standard cosmological model is defined by six parameters, which are taken to be $\left\lbrace \omega_\text{b}, \omega_\text{cdm}, \theta_\text{s}, \tau_\text{reio}, A_\text{s}, n_\text{s} \right\rbrace   $, if one assumes a PL primordial power spectrum like Eq. \eqref{PL-PS}. The first two parameters describe the densities of baryon and cold dark matter which are given by the parameters $ \omega_\text{b} = \Omega_\text{b} h^2 $ and $ \omega_\text{cdm} = \Omega_\text{cdm} h^2 $, where $ h $ is the dimensionless Hubble constant with a definition $ H_0 \equiv h \times 100\, \text{km}/\text{s}/\text{Mpc} $. $ \theta_\text{s} $ is the angular scale of the sound horizon. The angular wavenumber corresponding to the pivot scale is $ \ell_\ast \simeq \eta_0 k_\ast \sim 700 $. Here, $ \eta_0 $ is the conformal time today. At high-$\ell$ ($\ell > \ell_\ast $), due to the precise constraints on the anisotropy spectrum, a larger spectral index will reduce the parameter $ \omega_\text{cdm} $ and increase $ \omega_\text{b} $ slightly, as shown in Figure \ref{fig:LCDM_All}. The combination of these two effects will give a smaller matter density parameter $ \Omega_\text{m} $ in HZ-$\Lambda$CDM as $ \Omega_\text{m} = 0.2740 \pm 0.0040 $ than it in PL-$\Lambda$CDM  as $ \Omega_\text{m} = 0.3080 \pm 0.0062 $ with CMB data, see Table \ref{tab-LCDM}.   Since the gravitational potential is dominated by matter at recombination, changing the matter density will affect the angular scale of the sound horizon. Combining all the effects, a large spectral index will increase the derived parameter $ H_0 $, which gives the same conclusions as other methods to solve Hubble tension \cite{Li:2024rgq, Jiang:2022qlj, Wang:2024tjd, Jiang:2023bsz}.

\subsection{Structure growth parameter ($S_8$) tension}
\label{Sec:S8}

\begin{figure}[htbp]
	\begin{flushleft}
		\includegraphics[width=0.95\textwidth]{{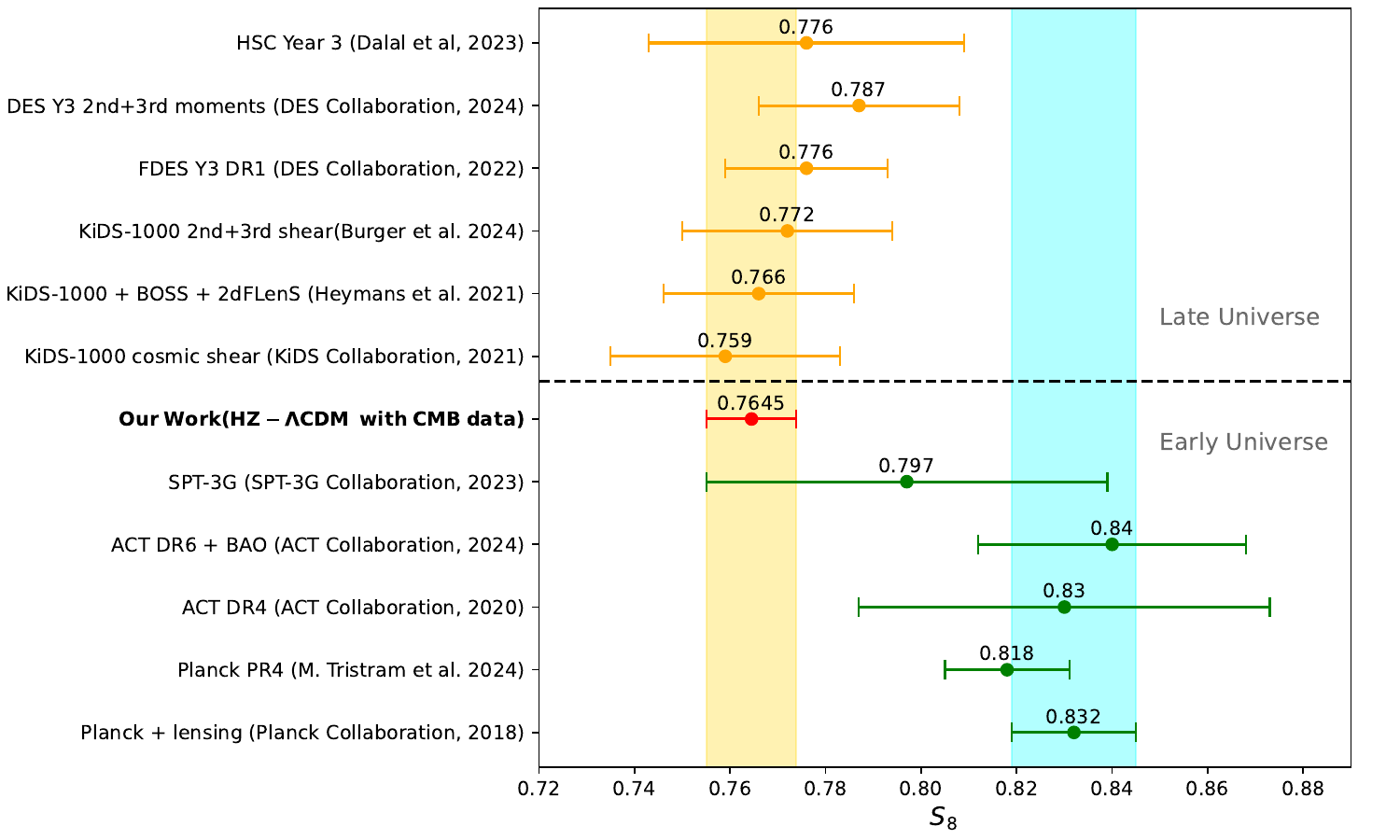}}
		\caption{\label{fig:S8} $S_8$ values of late-universe and early-universe observations. The cyan color band  presents the mean value and $ 1\sigma $ limit of \texttt{Planck 2018 TTTEEE}, \texttt{lowT}, \texttt{lowE} and \texttt{lensing data}, which is $ S_8 = 0.832 \pm 0.013 $\cite{Planck:2018vyg}. The gold color band is the mean value and $ 1\sigma $ limit of  HZ-$\Lambda$CDM model with CMB data, which is $ S_8 = 0.7645 \pm 0.0094 $ in our work. }
	\end{flushleft} 
\end{figure}

In addition to the Hubble tension, another cosmological tension named the $ S_8 $ tension exists from different measurements, where $ S_8 \equiv \sigma_8 \sqrt{\Omega_\text{m}/0.3} $ is the combined structure growth parameter.  $ \sigma_8 $ is the amplitude of matter fluctuations on the sphere of $8h^{-1}$Mpc. The CMB results under $ \Lambda $CDM show that $ S_8 = 0.834 \pm 0.016 $(\texttt{Planck 2018}) \cite{Planck:2018vyg}, $ S_8 = 0.797 \pm 0.042 $(\texttt{SPT-3G 2018}) \cite{SPT-3G:2022hvq} and $ S_8 = 0.840 \pm 0.028 $(\texttt{ACT DR6}+\texttt{BAO}) \cite{ACT:2023kun}. In contrast, dynamical cosmological probes favor lower values, which report $S_8 = 0.759^{+0.024}_{-0.021}$ (\texttt{KiDS-1000}) \cite{KiDS:2020suj}, $S_8 = 0.776 \pm 0.017$ (\texttt{DES-Y3}) \cite{DES:2021wwk}.  In addition to these, many local observations restrict the $ S_8 $ to a range $ 0.75 \sim 0.78 $ \cite{Dalal:2023olq,DES:2024jgw,Burger:2023qef, Heymans:2020gsg, DES:2021bvc, DES:2021vln}, which are in $2 \sim 3 \sigma$ tensions with the CMB results, as shown in Figure \ref{fig:S8}. A recent study with \texttt{DES-Y1} and the fourth data release of \texttt{KiDS-1000} combined obtains $ S_8 = 0.736^{+0.016}_{-0.018} $,  increasing the tension to $ 4.1 \sigma $ compared to Planck 2018 results \cite{Harnois-Deraps:2024ucb}.

In this work, as shown in Table \ref{tab-LCDM} and Figure \ref{fig:S8}, our analysis of the $ S_8 $ value with constraints from CMB data in the HZ-$ \Lambda\text{CDM}$ model is $ S_8 = 0.7645 \pm 0.0094 $, which is consistent with the results of local measurements. There is no tension of $S_8$ in HZ-$ \Lambda\text{CDM}$ model. However, in PL-$ \Lambda\text{CDM}$ model, $ S_8 = 0.823 \pm 0.012 $ with CMB data. There is still a nearly $ 3\sigma $ tension of $S_8$ between the value obtained from CMB in PL-$\Lambda$CDM model and values measured directly from \texttt{KiDS-1000} and \texttt{DES-Y3} \cite{Heymans:2020gsg, KiDS:2020suj,DES:2021wwk}. Since $ S_8 $ is a combined parameter to describe the growth of matter fluctuations, a smaller $ S_8 $ expresses a lower value of the matter density parameter. The $ \Omega_\text{m}$ in the HZ-$ \Lambda\text{CDM}$ model is smaller than its value in PL-$ \Lambda\text{CDM}$ model. In contrast to the Hubble constant, a larger spectral index produces a smaller value of $ S_8 $.

\section{Extensions of $ \Lambda $CDM: the negative cases to support Harrison-Zel'dovich spectrum }
\label{Sec:extensions}
\subsection{CMB lensing amplitude ($A_\text{L}$) anomaly}
\label{Sec:lensing}

Weak lensing of CMB can smooth the shape of the observed power spectra. The gravitational lensing amplitude is defined as a rescaling factor of lensing potential power spectrum \cite{Calabrese:2008rt}:
\begin{equation}
	C_{\ell}^{\phi\phi} \longrightarrow A_\text{L} C_{\ell}^{\phi\phi}.
\end{equation}
where $ A_\text{L} = 0 $ implies that the CMB is unlensed, while $ A_\text{L} = 1 $ is the expected value in $ \Lambda $CDM model. There is a nearly $ 3\sigma $  deviation from PL-$\Lambda$CDM$ + A_\text{L}$  in the Planck 2018 result with $ A_\text{L} = 1.18\pm 0.065 \left( 68\%, \, \texttt{Planck2018 TT,TE,EE+lowE} \right)  $ \cite{Planck:2018vyg}, implying a CMB lensing anomaly.

\begin{figure}[htbp]
	\centering
        \includegraphics[width=1.0\textwidth]{{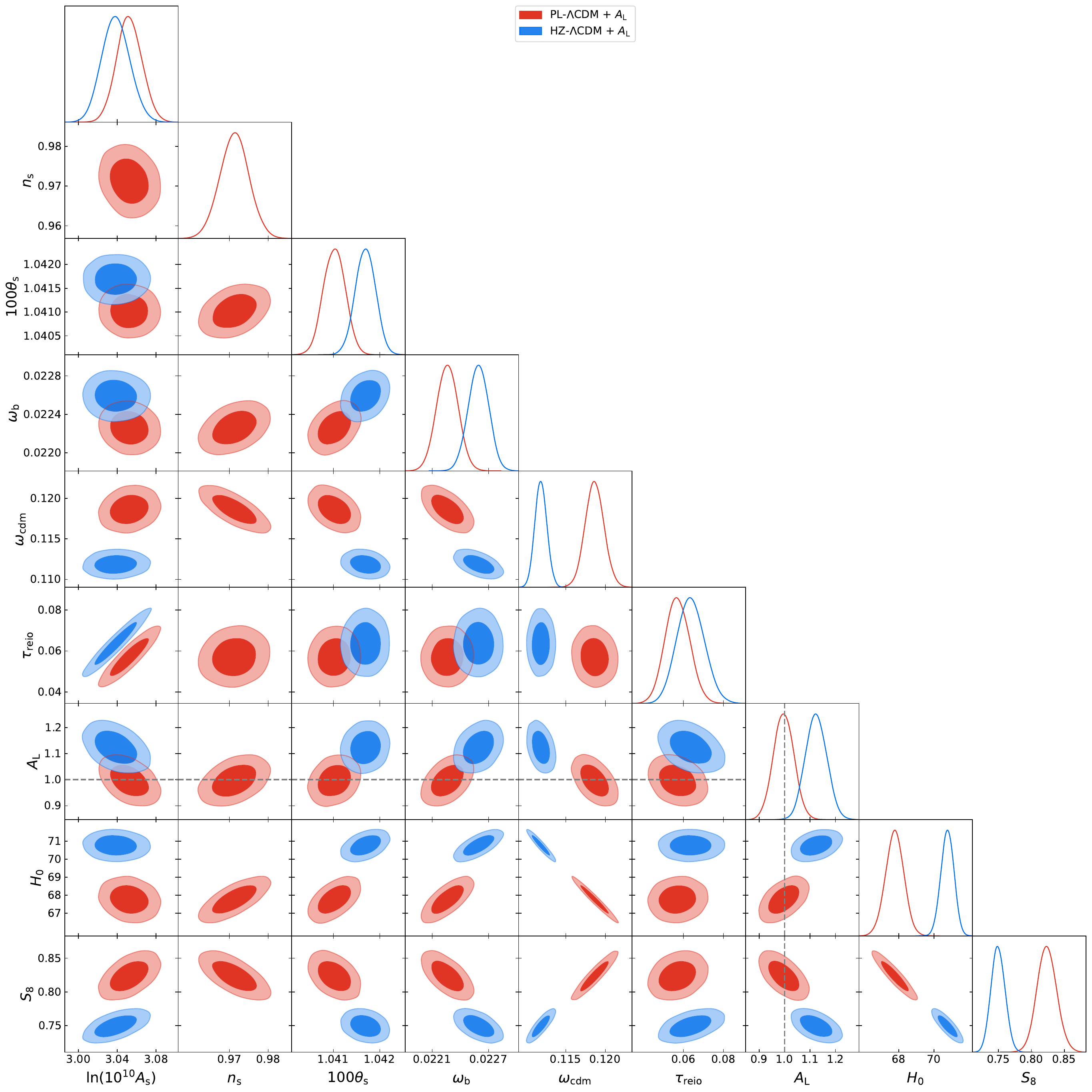}}
		\caption{\label{fig:LCDM_Al_All} Triangle plot for the cosmological parameters with CMB data in PL/HZ-$\Lambda$CDM$ + A_\text{L}$ model with CMB data. The dashed line marks $ A_\text{L} =1 $ which is the expected value in $ \Lambda $CDM model.}
\end{figure}

We consider the simplest extension of $ \Lambda $CDM by adding the lensing amplitude $A_\text{L}$.  As shown in Table \ref{tab-LCDM_Al} and Figure \ref{fig:LCDM_Al_All}, in this PL-$\Lambda$CDM$ + A_\text{L}$ model with CMB data, $A_\text{L} = 0.997 \pm 0.040 $ that have no anomaly compared with the prediction by the standard cosmological model. In HZ-$\Lambda$CDM$ + A_\text{L}$ model, although the value of the lensing amplitude is smaller than the Planck 2018 result \cite{Planck:2018vyg}, that $A_\text{L} = 1.124\pm 0.042  $, there is still about $ 3 \sigma $ difference with unity. In other words,  HZ spectrum will aggravate the lensing anomaly.

\begin{table}[htbp]
	\centering
	\begin{tabular} {| l | c | c | }
	\hline
	Parameter &  PL-$\Lambda$CDM$ + A_\text{L}$   &  HZ-$\Lambda$CDM$ + A_\text{L}$\\
\hline
{$A_\text{L}     $} & $0.997\pm 0.040            $ & $1.124\pm 0.042$\\

{$\ln(10^{10} A_\text{s})$} & $3.052\pm 0.013            $ & $3.039\pm 0.014$\\

{$n_\text{s}     $} & $0.9713\pm 0.0038          $ & \\

{$100\theta_\text{s}$} & $1.04102\pm 0.00023        $ & $1.04169\pm 0.00022$\\

{$\omega_\text{b}$} & $0.02226\pm 0.00011        $ & $0.02259\pm 0.00011$\\

{$\omega_\text{cdm}$} & $0.1186\pm 0.0012          $ & $0.11185\pm 0.00075$\\

{$\tau_\text{reio}$} & $0.0571\pm 0.0061          $ & $0.0637\pm 0.0069$\\
\hline
$H_\text{0}                $ & $67.77\pm 0.52             $ & $70.77\pm 0.37$\\

$\Omega_\text{m}           $ & $0.3083\pm 0.0072          $ & $0.2698\pm 0.0041$\\

$z_\text{reio}             $ & $7.96\pm 0.61              $ & $8.40\pm 0.65$\\

$S_\text{8}                $ & $0.824\pm 0.015            $ & $0.749\pm 0.011$\\
\hline
\end{tabular}
	\caption{\label{tab-LCDM_Al} Cosmological parameters with mean value and $ 68\% $ limits by using CMB data in PL/HZ-$\Lambda$CDM$ + A_\text{L}$ model. The lensing amplitude $A_\text{L}$ should be unity, as predicted by the $\Lambda$CDM universe.} 
\end{table}

 In recent data release, the lensing anomaly is no longer significant. In PL-$\Lambda$CDM model, the lensing amplitude of different date sets are $A_{\text{L}}=1.039\pm0.052$ (68\%, \texttt{Planck PR4}), $A_{\text{L}}=0.87\pm0.11$ (68\%, \texttt{SPT3G}) \cite{SPT-3G:2022hvq} and $A_{\text{L}}=1.01\pm0.11$ (68\%, \texttt{ACT DR4}) \cite{ACT:2020gnv}. The $A_\text{L}$ measured from \texttt{Planck PR4} and \texttt{ACT DR4} are a little lager than unity but under $1 \sigma$ level. The result in \texttt{SPT3G} data interprets a lower value of $A_\text{L}$ than unity and shows a $1.2 \sigma$ difference from unity. Thus, our result that $A_\text{L} = 0.997 \pm 0.040 $ is close to unity in PL-$\Lambda$CDM, as expected when combined with these data sets.

Fortunately, the cosmological tensions are alleviated in the extended model with HZ spectrum, which gives $ H_\text{0} = 70.77\pm 0.37  $ and $ S_8 = 0.749\pm 0.011 $. The Hubble constant is a little larger than its value in HZ-$\Lambda$CDM with CMB data but is still consistent with the result of CCHP group. Similarly, the structure growth parameter is a little smaller than its value in HZ-$\Lambda$CDM but is still consistent with the late observations \cite{Heymans:2020gsg,KiDS:2020suj,Harnois-Deraps:2024ucb}.

Another notable point is that HZ-$\Lambda$CDM$ + A_\text{L}$ model yields a relatively smaller optical depth/redshift of reionization compared with HZ-$\Lambda$CDM, that is $ \tau_\text{reio} = 0.0637\pm 0.0069$ ($ z_\text{reio} = 8.40\pm 0.65 $). The EoR described by JWST that $ z_\text{reio,JWST} \approx 8.9 $ is still compatible with this case at $ 1\sigma $ level.  The suppression on CMB anisotropies from reionization, as a factor of $e^{-2\tau_{\text{reio}}}$, introduces a strong degeneracy between optical depth $ \tau_\text{reio} $ and gravitational lensing amplitude $ A_\text{L} $, coming with the parameters \{$ A_\text{s} $, $ n_\text{s} $\}, as shown in Figure \ref{fig:LCDM_Al_All}.

\subsection{Spatial curvature density ($\Omega_\text{k}$) anomaly }
\label{Sec:curvature}

Pure Planck data prefer a negative value of the spatial curvature density parameter , like $ \Omega_\text{k}=-0.0106\pm0.0065 $ (68\%, \texttt{Planck2018 TT,TE,EE+lowE+lesing}), in PL-$\Lambda$CDM$+ \Omega_\text{k} $ model \cite{Planck:2018vyg,Tristram:2023haj,DiValentino:2019qzk,Rosenberg:2022sdy}, which implies that our universe is closed. This difference with the prediction from $\Lambda$CDM that $ \Omega_\text{k} $ should be zero is named spatial curvature density anomaly. However, when BAO data is added to break the geometrical degeneracy, the curvature density $ \Omega_\text{k} $ usually goes to zero, or even becomes slightly greater than zero, like $\Omega_{\text{k}}=0.0007\pm0.0019$ (68\%, \texttt{Planck2018 TT,TE,EE+lowE+lesing+BAO}), but still including $ \Omega_\text{k} =0 $ at $ 1 \sigma$ level \cite{Planck:2018vyg,Tristram:2023haj,DESI:2024mwx}. Thus the universe is flat or even prefer to become open when some date sets like BAO and conservative Big Bang Nucleosynthesis (BBN) priors are considered \cite{DESI:2024mwx}.

\begin{figure}[htbp]
	\begin{flushleft}
		\includegraphics[width=1.0\textwidth]{{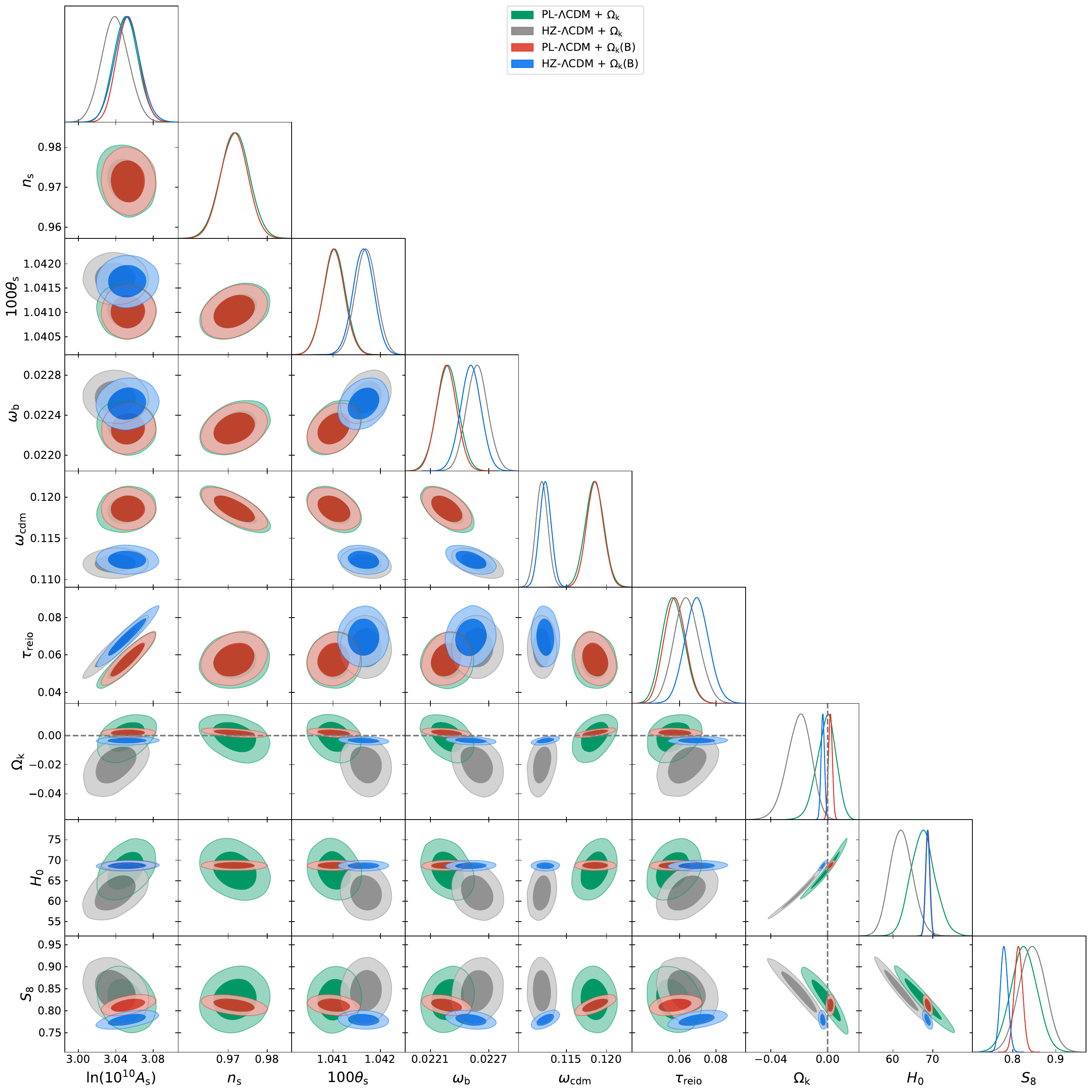}}
		\caption{\label{fig:LCDM_BAO_k_All} Triangle plot for cosmological parameters with CMB data or CMB $+$ DESI BAO data in PL/HZ-$\Lambda$CDM$+ \Omega_\text{k}$ model. The dashed line marks $ \Omega_\text{k} =0 $ which is the expected value in flat-$ \Lambda $CDM model. The mark ``(B)'' also means adding BAO data. The parameter space is constrained to be more accurate if the DESI BAO data is added.} 
	\end{flushleft} 
\end{figure}

In this work, we consider both the CMB data and CMB $+$ BAO data, see Table \ref{tab-LCDM_k} and Figure \ref{fig:LCDM_BAO_k_All}. The curvature density of HZ-$\Lambda$CDM$+ \Omega_\text{k} $ model is more negative than its value of PL-$\Lambda$CDM$+ \Omega_\text{k} $ model in both of the two date sets. In PL-$\Lambda$CDM$+ \Omega_\text{k} $ model with only CMB data, $\Omega_\text{k} = -0.0011^{+0.0074}_{-0.0061}$, which includes zero  at $ 1\sigma $ level, but there is still about $ 2.2 \sigma$ away from zero in HZ-$\Lambda$CDM$+ \Omega_\text{k} $ model that $\Omega_\text{k} = -0.0204^{+0.0091}_{-0.0075}$ with only CMB data. Adding the \texttt{DESI BAO} data (marked by ``(B)'' in Table \ref{tab-LCDM_k} and Figure \ref{fig:LCDM_BAO_k_All}), the parameter space of curvature density $ \Omega_\text{k} $ is constrained more accurately. In  PL-$\Lambda$CDM$+ \Omega_\text{k} $(B) case, $\Omega_\text{k}=0.0019\pm 0.0014 $  including zero at $1.3 \sigma$ level, which gives a preference to an open universe. Like lensing anomaly, the result of spatial curvature density is still negative in HZ-$\Lambda$CDM$+ \Omega_\text{k} $(B) case.  Although $ \Omega_\text{k} = -0.0035\pm 0.0012 $ in HZ-$\Lambda$CDM+$\Omega_{\text{k}}$(B) case is closer to zero than HZ-$\Lambda$CDM$+ \Omega_\text{k} $ case with only CMB data, it is still negative and there is a nearly $ 3 \sigma $ deviation from zero when adding \texttt{DESI BAO} data.

\begin{table}[htbp]
	\centering
	\begin{tabular} {| l | c | c | c | c |}
		\hline
		Parameter &  PL-$\Lambda$CDM$+ \Omega_\text{k}$  &  HZ-$\Lambda$CDM$+ \Omega_\text{k}$ &  PL-$\Lambda$CDM$+ \Omega_\text{k}$(B) &  HZ-$\Lambda$CDM$+ \Omega_\text{k}$(B)\\
\hline
{$\Omega_\text{k}       $} & $-0.0011^{+0.0074}_{-0.0061}$ & $-0.0204^{+0.0091}_{-0.0075}$ & $0.0019\pm 0.0014          $ & $-0.0035\pm 0.0012         $\\

{$\ln(10^{10} A_\text{s})$} & $3.051\pm 0.013            $ & $3.039\pm 0.014            $ & $3.053\pm 0.012            $ & $3.052\pm 0.014            $\\

{$n_\text{s}            $} & $0.9717\pm 0.0037          $ &                              & $0.9715\pm 0.0035          $ &                             \\

{$100\theta_\text{s}  $} & $1.04103\pm 0.00023        $ & $1.04169\pm 0.00022        $ & $1.04102\pm 0.00023        $ & $1.04164\pm 0.00022        $\\

{$\omega_\text{b}       $} & $0.02228\pm 0.00011        $ & $0.02258\pm 0.00011        $ & $0.02227\pm 0.00010        $ & $0.02251\pm 0.00011        $\\

{$\omega_\text{cdm}   $} & $0.1185\pm 0.0012          $ & $0.11195\pm 0.00075        $ & $0.1186\pm 0.0011          $ & $0.11237\pm 0.00073        $\\

{$\tau_\text{reio}    $} & $0.0566\pm 0.0062          $ & $0.0639\pm 0.0069          $ & $0.0576\pm 0.0059          $ & $0.0696\pm 0.0066          $\\
\hline
$H_\text{0}                       $ & $67.5^{+2.8}_{-3.3}        $ & $62.1^{+2.7}_{-3.0}        $ & $68.74\pm 0.53             $ & $68.67\pm 0.53             $\\

$\Omega_\text{m}                  $ & $0.312^{+0.026}_{-0.029}   $ & $0.353^{+0.030}_{-0.034}   $ & $0.2995\pm 0.0050          $ & $0.2875\pm 0.0047          $\\

$z_\text{reio}                  $ & $7.90\pm 0.63              $ & $8.35\pm 0.67              $ & $8.02\pm 0.59              $ & $8.97\pm 0.61              $\\

$S_\text{8}                       $ & $0.826\pm 0.031            $ & $0.845\pm 0.031            $ & $0.8129\pm 0.0097          $ & $0.7795\pm 0.0088          $\\
\hline
	\end{tabular}
	\caption{\label{tab-LCDM_k} Cosmological parameters with mean value and $ 68\% $ limits by using CMB data or CMB $+$ DESI BAO data in PL/HZ-$\Lambda$CDM$+ \Omega_\text{k}$ model. The mark ``(B)'' means adding BAO data. }
\end{table}

In $\Lambda$CDM$+ \Omega_\text{k}$ universe, since $ \Omega_\text{k} + \Omega_\text{m} + \Omega_\Lambda = 1 $, changing $\Omega_\text{k}$ will affect the matter density parameter and the value of cosmological constant. These parameters determine the gravitational potential and the angular diameter distance to last scattering, which directly affect the derived parameters $ H_0 $ and $S_8$. The curvature density is positively correlated with the Hubble constant, and negatively correlated with the structure growth parameter, see Figure \ref{fig:LCDM_BAO_k_All}. Thus increasing the $ \Omega_\text{k} $ is helpful to solve these two cosmological tensions. In PL-$\Lambda$CDM$+ \Omega_\text{k}$ model, adding BAO data, the curvature density becomes larger, even is positive, but the Hubble constant, the optical depth, and the structure growth parameter increase or decrease slightly. On the other hand, in the HZ-$\Lambda$CDM$+ \Omega_\text{k}$ model, adding BAO data, the curvature density will increase but is still negative. However, the Hubble constant, the optical depth, and the structure growth parameter can increase or decrease enough.  The curvature density parameter constrained by CMB $+$ BAO data in the PL-$\Lambda$CDM$+ \Omega_\text{k}$ model is greater than its value in the HZ-$\Lambda$CDM$+ \Omega_\text{k}$ model, but the Hubble constant is almost equal in these two models, as shown in the third and fourth columns of Table \ref{tab-LCDM_k}. Also, in the PL-$\Lambda$CDM$+ \Omega_\text{k}$ model, the optical depth is smaller and the structure growth parameter is larger compared to the values in the HZ-$\Lambda$CDM$+ \Omega_\text{k}$ model. These results show that the red tilt of the primordial power spectrum will exacerbate the JWST optical depth puzzle and cosmological tensions.

It is still hard to determine whether our universe is closed, flat, or open. The CMB constraints on the spatial curvature density are partially degenerate with those on the matter density and the cosmological constant if no large-scale structure probes like BAO or BBN priors are included to break the degeneracy. The parameter spaces of $\Omega_\text{k}$ (including two derived parameters $H_0$ and $ S_8 $) are large in this extended model by using only CMB data, both in HZ-$\Lambda$CDM and PL-$\Lambda$CDM, as shown in the second and third columns in Table \ref{tab-LCDM_k}, also see Figure \ref{fig:LCDM_BAO_k_All}. Moreover, from $\Omega_{\text{k}}=-0.0011^{+0.0074}_{-0.0061}$ in PL-$\Lambda$CDM$+\Omega_{\text{k}}$ model to $\Omega_{\text{k}}=-0.0204^{+0.0091}_{-0.0075}$ in HZ-$\Lambda$CDM$+\Omega_{\text{k}}$ model, it shows that the HZ primordial power spectrum tends to give a more negative value of $ \Omega_\text{k} $ that aggravate the spatial curvature density anomaly to about 2.2$\sigma$. And the HZ spectrum make cosmological tensions worse, like from $H_0=67.5^{+2.8}_{-3.3}$, $S_8 = 0.826\pm 0.031 $ in PL-$\Lambda$CDM$+ \Omega_{\text{k}}$ to $H_0=62.1^{+2.7}_{-3.0}$, $S_8 = 0.845\pm 0.031 $ in HZ-$\Lambda$CDM$+ \Omega_{\text{k}}$.  However, adding the BAO data resolves theses pessimism, both in PL-$\Lambda$CDM$+\Omega_{\text{k}}$ and HZ-$\Lambda$CDM$+\Omega_{\text{k}}$. Breaking geometrical degeneracy by BAO data usually prefers a value of $ \Omega_\text{k} $ close to zero, especially in the HZ-$\Lambda$CDM$+ \Omega_\text{k}$(B) model that $\Omega_{\text{k}}=-0.0035\pm0.0012$ with CMB $+$ BAO data, which can be acceptable at the level $ \lvert \Omega_\text{k} \rvert \sim \mathcal{O}(10^{-3}) $. The cosmological tensions can also be alleviated by adding BAO data. The Hubble values become $H_0 = 68.74\pm 0.53  $ in PL-$\Lambda$CDM$+ \Omega_\text{k}$(B) model and  $ H_0 = 68.67\pm 0.53  $ in HZ-$\Lambda$CDM$+ \Omega_\text{k}$(B) model, although both of them are still inconsistent with local observations at $4\sigma$ level. The $S_8 = 0.8129\pm 0.0097 $ in PL-$\Lambda$CDM$+ \Omega_\text{k}$(B) model still shows a $2 \sim 3 \sigma$ tension, but $S_8 = 0.7795\pm 0.0088 $ in HZ-$\Lambda$CDM$+ \Omega_\text{k}$(B) model have no tension with local observations at $1\sigma $ level.

\section{Discussion: the power-law spectrum, lensing amplitude and optical depth}
\label{Sec:discussion}
\subsection{Suppressing the large-scale CMB angular power spectrum}

The CMB angular power spectrum at low multipoles (low-$\ell$ or large scales) observed by COBE, WMAP and Planck satellites appears to be lower than the one predicted by $\Lambda$CDM model with a PL form of primordial power spectrum \cite{Hinshaw:1996ut,WMAP:2003ivt,Planck:2013lks,Planck:2019evm}. This lack of large-angle CMB temperature correlations is also one of the CMB anomalies. There are some statistical treatments and potential explanations to explain this lack of correlations, like the $S_\text{1/2}$ statistic first introduced in the WMAP first-year release \cite{WMAP:2003elm}, the ISW effect \cite{Copi:2016hhq} and others \cite{Planck:2019evm,Gruppuso:2013dba,Aurich:2021ofm,Bernui:2018wef,Pranav:2018lox}. Besides these statistical views, previous works show that the Galilean genesis, string gas cosmology and bounce scenario can also explain the suppressing power spectrum of CMB temperature anisotropies at large angular scales \cite{Li:2024rgq, Creminelli:2010ba, Brandenberger:2008nx, Piao:2003zm, Liu:2013kea, Qiu:2015nha,Ni:2017jxw}.

\begin{figure}[htbp]
	\centering
	\includegraphics[width=0.98\textwidth]{{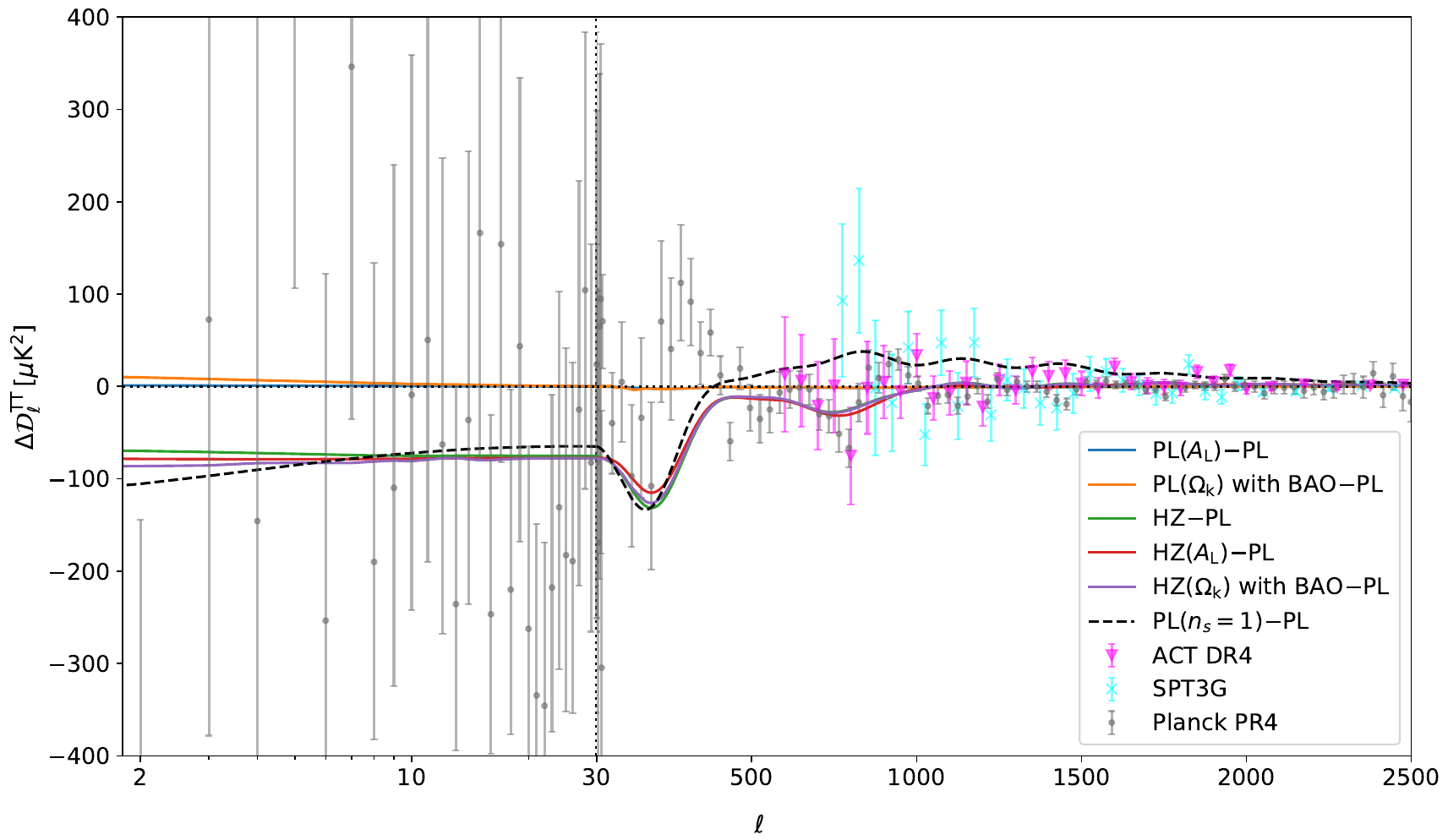}}
	\caption{\label{fig:DlTT_Delta} Deviations of CMB temperature angular power spectra from the fiducial PL-$\Lambda$CDM model. The legend `` PL '' represents the PL-$\Lambda$CDM model, and The legend `` HZ '' represents the HZ-$\Lambda$CDM model.  The `` $A_\text{L}$ '' and `` $\Omega_\text{k}$ '' in brackets present the extensions. Residuals with respect to PL-$\Lambda$CDM model. The parameters we used are the best-fit values from Table \ref{tab-LCDM}, Table \ref{tab-LCDM_Al} and Table \ref{tab-LCDM_k}. Parameters of PL are from the second column in Table \ref{tab-LCDM}. The legend ``PL($n_\text{s} = 1$)'' of the dashed black line means that we use the parameters from the second column in Table \ref{tab-LCDM} and only set the spectral index $ n_\text{s} = 1 $ artificially. The non-smoothness at $ \ell = 30 $ is due to the logarithmic coordinates at $ \ell < 30 $ and linear coordinates at $ \ell > 30 $. }
\end{figure}

CMB observations show a red tilt of the PL spectrum parameterized by equation \eqref{PL-PS}. Unlike the genesis or bounce scenarios, which only give the lower amplitudes of the primordial power spectrum at large scales and barely change the spectrum at small scales, in the HZ spectrum case, the amplitude is lower than one of the PL spectrum with a red tilt at small $k$ ($k < k_\ast$) and higher at large $k$ ($k > k_\ast$). That means the CMB angular power spectra can be suppressed at low $\ell$ ($\ell < \ell_\ast \sim 700 $) and can be uplifted at high $\ell$ ($\ell > \ell_\ast$). We plot the deviation of angular power spectra of the CMB temperature fluctuations from the fiducial PL-$\Lambda$CDM model to show this effect, see Figure \ref{fig:DlTT_Delta}. 

This suppressing effect at low $\ell$ may explain the lack of large-angle CMB temperature correlations. In Figure \ref{fig:DlTT_Delta}, the orange and blue lines show almost no deviation of angular power spectra, because $ A_\text{L} \cong 1 $ and $ \Omega_\text{k} \sim 0 $ in the extended cases are closed to the prediction of the fiducial PL-$\Lambda$CDM model. In the HZ spectrum case, all the three lines show some obviously suppressions at low $\ell$ compared with the fiducial PL-$\Lambda$CDM model. The large cosmic variance errors at low $\ell$ allow this suppressing, but at high $\ell$, due to the angular power spectrum is detected precisely, the uplifted effects from HZ spectrum should be offset by other approaches, like some smoothing effects from gravitational lensing, reionization or changing the density of baryon and CDM.

\subsection{Smoothing effects of CMB angular power spectrum}

In the PL spectrum cases,  if we artificially change the spectral index from its best-fit value $ n_\text{s} = 0.9716 \pm 0.0035 $ in Table \ref{tab-LCDM} to $ n_\text{s} = 1 $ (not change other fiducial parameters in the second column in Table \ref{tab-LCDM}), the angular power spectrum is also suppressed at low $\ell $, like the HZ spectrum. However, the angular power spectrum is enhanced much at high $\ell$, and shows a large deviation from other cases and residuals, see the dashed black line in Figure \ref{fig:DlTT_Delta}. Since in the HZ spectrum cases, other cosmological parameters are free, the enhancements at high $\ell$ are compensated by changing other parameters. For example, the lower $A_\text{s}$ and $\omega_\text{cdm}$, the larger $\theta_\text{s}$, $\omega_\text{b}$ and $ \tau_\text{reio} $, see Table \ref{tab-LCDM} and Figure \ref{fig:LCDM_All}.

\begin{figure}[htbp]
	\centering
	\includegraphics[width=0.98\textwidth]{{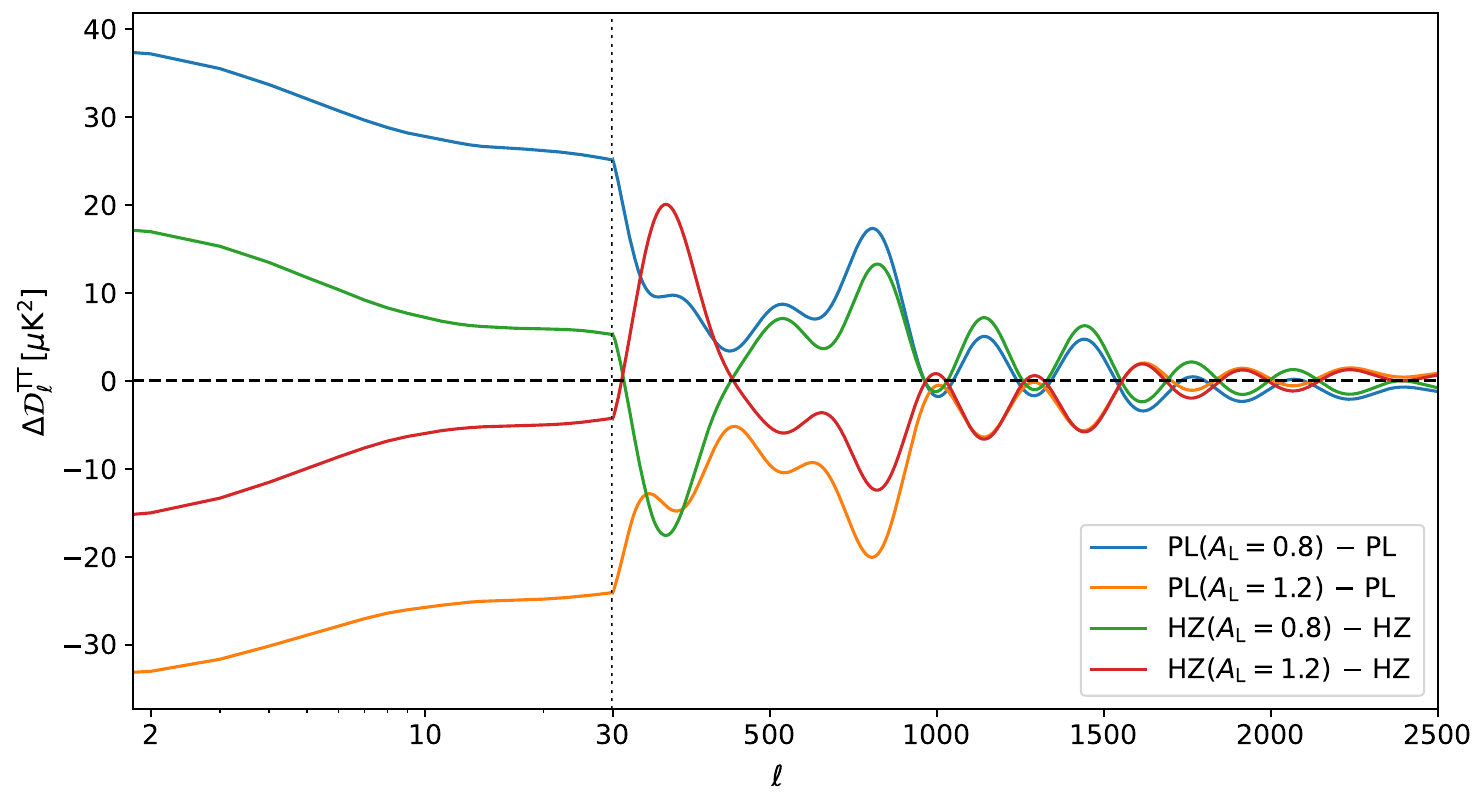}}
	\caption{\label{fig:DlTT_Delta_HZAl} Deviations of the CMB temperature angular power spectra from the fiducial PL-$\Lambda$CDM model. $ A_\text{L} $ are fixed as $0.8$  or $1.2$ to see the effect on the angular power spectrum in this analysis. The parameters we used are the best-fit values from Table \ref{table:Al-fixed}. We use logarithmic coordinates at $ \ell < 30 $ and linear coordinates at $ \ell > 30 $. }
\end{figure}

As mentioned before, gravitational lensing and reionization of the universe will smooth the CMB angular power spectrum and wash out the primordial anisotropies.  The CMB power spectrum is suppressed from reionization by a factor of $ e^{-2\tau_\text{reio}}$ corresponding to multipoles $ \ell > \eta_0/\eta_\text{reio} \sim 10 $, since the effect only occurs on scales that are smaller than the horizon at reionization.  When fixing other cosmological parameters and only changing the value of $ A_\text{L} $, a larger $ A_\text{L} $ will suppress peaks and uplift valleys of the CMB temperature angular power spectrum \cite{Calabrese:2008rt}. But when we fix $ A_\text{L} $, and fit other cosmological parameters, the results become more complicated, as shown in Figure \ref{fig:DlTT_Delta_HZAl} and Table \ref{table:Al-fixed}. 

\begin{table}
\centering
\begin{tabular} {| l | c |c |c| c|}
\hline
 Parameter &   \multicolumn{2}{|c|}{PL-$\Lambda$CDM} &   \multicolumn{2}{|c|}{HZ-$\Lambda$CDM}\\
 \hline
     & $A_\text{L} = 0.8$ & $A_\text{L} = 1.2$ & $A_\text{L} = 0.8$ & $A_\text{L} = 1.2$\\

\hline
{\boldmath$\log(10^{10} A_\text{s})$} & $3.077\pm 0.012            $ & $3.026\pm 0.011            $ & $3.083\pm 0.015            $ & $3.028\pm 0.012            $\\

{\boldmath$n_\text{s}     $} & $0.9641\pm 0.0036          $ & $0.9788\pm 0.0033          $ &                              &                             \\

{\boldmath$100\theta_\text{s}$} & $1.04078\pm 0.00023        $ & $1.04129\pm 0.00023        $ & $1.04151\pm 0.00021        $ & $1.04173\pm 0.00021        $\\

{\boldmath$\omega_\text{b}$} & $0.02202\pm 0.00011        $ & $0.02249\pm 0.00010        $ & $0.02232\pm 0.00010        $ & $0.02265\pm 0.00010        $\\

{\boldmath$\omega_\text{cdm}$} & $0.1213\pm 0.0011          $ & $0.11577\pm 0.00098        $ & $0.11364\pm 0.00074        $ & $0.11141\pm 0.00068        $\\

{\boldmath$\tau_\text{reio}$} & $0.0644\pm 0.0062          $ & $0.0493\pm 0.0056          $ & $0.0829\pm 0.0073          $ & $0.0591\pm 0.0061          $\\

\hline

$H_\text{0}                $ & $66.54\pm 0.47             $ & $69.08\pm 0.44             $ & $69.80\pm 0.35             $ & $71.00\pm 0.33             $\\

$\Omega_\text{m}           $ & $0.3253\pm 0.0069          $ & $0.2911\pm 0.0056          $ & $0.2804\pm 0.0041          $ & $0.2673\pm 0.0037          $\\

$z_\text{reio}             $ & $8.78\pm 0.60              $ & $7.07\pm 0.57              $ & $10.26\pm 0.64             $ & $7.94\pm 0.58              $\\

$S_\text{8}                $ & $0.864\pm 0.013            $ & $0.781\pm 0.011            $ & $0.789\pm 0.010            $ & $0.7401\pm 0.0086          $\\
\hline
\end{tabular}
\caption{\label{table:Al-fixed} Cosmological parameters with mean value and 68\% limits by using CMB data in PL/HZ-$\Lambda$CDM$+ A_\text{L}$ model. But the parameter $ A_\text{L} $ are fixed as $0.8$ or $1.2$. }
\end{table}

We set the lensing amplitude $ A_\text{L} $ as two fixed values to see the changes of CMB angular power spectra in this analysis, which are $0.8$ and $1.2$. One of them is smaller than $1$ and another is larger than $1$.  When $ A_\text{L} = 0.8 $, both PL and HZ spectra tend to give an uplifted angular power spectrum at low $ \ell $. At high $ \ell $, this small lensing amplitude can uplift  peaks\footnote{Notice that the peak values of temperature angular power spectrum at high-$\ell$ is 4-th peak at $\ell \approx 1120$, 5-th peak at $\ell \approx 1445$ and so on \cite{Pan:2016zla}.}  and suppress valleys of these angular power spectra. When $  A_\text{L} = 1.2  $, we get opposite results. The same conclusion as \cite{Calabrese:2008rt} only occurs at high-$ \ell $ part. Regardless of whether $  A_\text{L} $ is greater or less than 1, the relative deviations in the HZ spectrum model are smaller than those in the case of a PL spectrum.

Comparing the results in HZ-$\Lambda$CDM model ($A_\text{L} = 1$ by default) and the extended HZ-$\Lambda$CDM$+ A_\text{L}$ model ($A_\text{L} = 1.124 \pm 0.042$ by fitting), a larger lensing amplitude favors a cosmology with the lower $A_\text{s}$, $ \omega_\text{cdm} $ and $ \tau_\text{reio} $, the larger $\theta_\text{s}$, $\omega_\text{b}$, see Table \ref{tab-LCDM} and Table \ref{tab-LCDM_Al} (also see Table \ref{table:Al-fixed}). When free the $A_\text{L}$ in HZ-$\Lambda$CDM, some part of the smoothing effect from optical depth can be replaced by CMB lensing, thus the $A_\text{L}$ will increase and optical depth will decrease in HZ-$\Lambda$CDM$+ A_\text{L}$ model compared with HZ-$\Lambda$CDM model.  Since the acoustic feature of the CMB angular power spectrum tightly constraints the parameter combination $ \Omega_\text{m}h^3 $\cite{Murgia:2020ryi}. A lower $\omega_\text{cdm}$ will cause a higher $H_0$ and a smaller $S_8$. The Hubble parameter $H_0$ and structure growth parameter $S_8$ change from $H_0 = 70.38 \pm 0.35$, $ S_8 = 0.7645 \pm 0.0094 $ in HZ-$\Lambda$CDM model to $H_0 = 70.77 \pm 0.37$, $ S_8 = 0.749 \pm 0.011 $ in HZ-$\Lambda$CDM$+ A_\text{L}$ model, although both of the changes are less than $ 1\sigma $ significant. 

\begin{figure}[htbp]
		\centering
        \includegraphics[width=0.98\textwidth]{{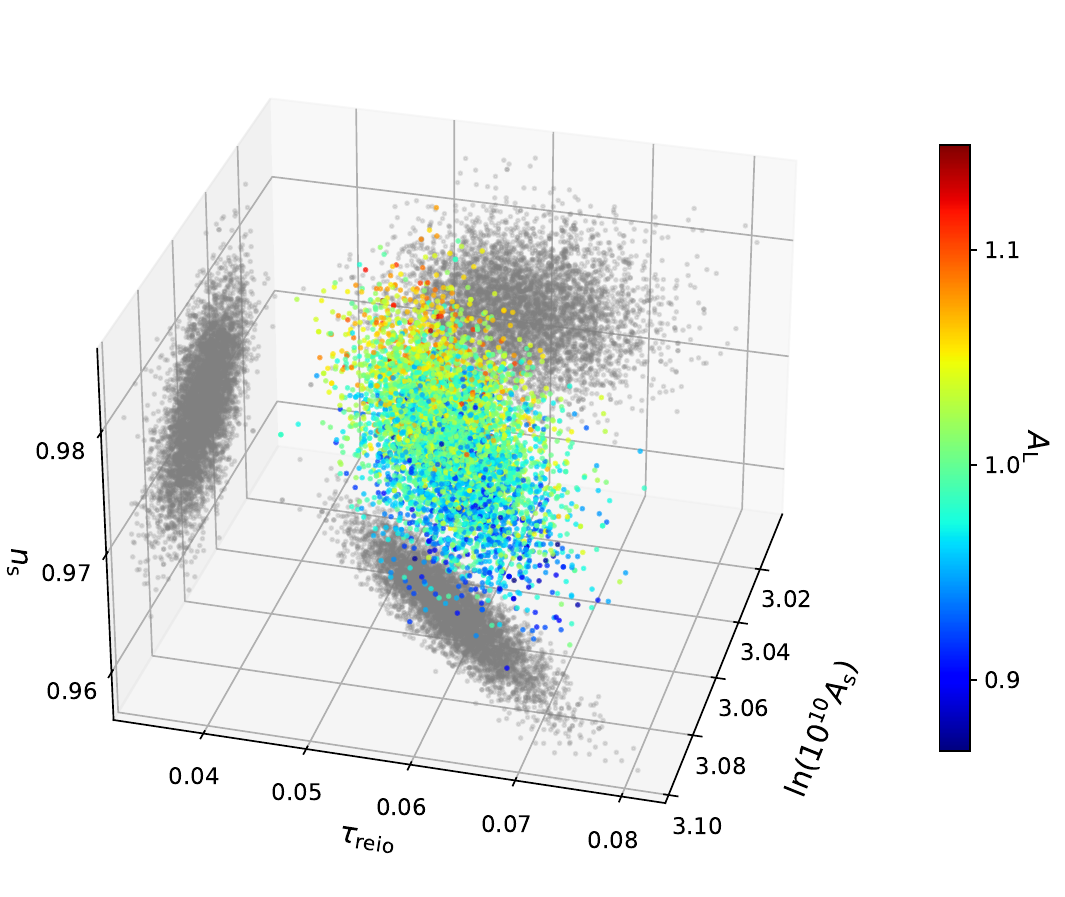}}
		\caption{\label{fig:As_tau_ns_Al} 3D plot for some cosmological parameters with CMB data in PL-$\Lambda$CDM$+ A_\text{L}$ model. When  the $ n_\text{s} $ becomes larger, the $ \tau_\text{reio} $ and the $A_\text{L}$ will increase but the $ A_\text{s} $ will reduce. } 
\end{figure}

The smoothing effect can also be mimicked by the primordial power spectrum combined with two parameters \{$ A_\text{s}, n_\text{s} $\}. As shown in Figure \ref{fig:DlTT_Delta_HZAl}, some uplifting/suppressing effects of weak lensing are compensated by using the HZ spectrum, especially at low multipoles. We also show that there are some strong degeneracies among the optical depth $ \tau_\text{reio} $,  the gravitational lensing amplitude $ A_\text{L} $ and parameters \{$ A_\text{s}, n_\text{s} $\}, as shown in the 3D plotted Figure \ref{fig:As_tau_ns_Al}. A larger spectral index will increase the optical depth and reduce the lensing amplitude to offset an uplift at high $\ell$ and a suppression at low $\ell$ of the angular power spectrum.  As the result from article \cite{Murgia:2020ryi}, there are other smoothing components which cannot be attributed to actual gravitational lensing. Until now, it seems that we can't distinguish these smoothing effects clearly. 

Since the optical depth of reionization is degenerate with the amplitude of the primordial power spectrum, observations usually constrain a combination parameter $ A_\text{s} e^{-2 \tau_{\text{reio}}}$. Some independent measurements are needed to break this degeneracy.  The CMB polarizations provide a chance to detect the optical depth accurately. In this work, we show that local observations form JWST provide another opportunity to restrict the optical depth. Previous works claim that the spatial curvature density anomaly may be a plausible source of the anomalous lensing amplitude \cite{DiValentino:2019qzk,DiValentino:2020srs,Handley:2019tkm}. Considering the degeneracies among those parameters, new methods are needed to measure the lensing amplitude  precisely. Maybe there are no reionization puzzles and CMB anomalies, such as lensing or curvature density anomaly, when the cosmological parameters can be constrained independently.

\section{Conclusion}
\label{conclusion}

 The redshift of reionization obtained from CMB estimation is less than the observation of JWST, which implies a reionization puzzle \cite{Munoz:2024fas}. This puzzle also appears in our analysis of combination with \texttt{Planck PR4}, \texttt{SPT3G}, and \texttt{ACT DR4} data sets with PL-$\Lambda$CDM cosmology, and we show that it can be avoided by choosing the HZ spectrum as a different initical condition. A larger spectral index needs the increasing of the optical depth or the redshift of reionization. That is to say, the EoR is earlier in HZ-$\Lambda$CDM model than previously expected.

 Furthermore, $H_0=70.38\pm0.35 \, \text{km}/\text{s}/\text{Mpc}$ in HZ-$\Lambda$CDM, leaving about 2.5$\sigma$ deviation compared with local observations $ H_0 = 73.04 \pm 1.04 \, \text{km}/\text{s}/\text{Mpc}$ using the Cepheid-SN Ia sample from HST  \cite{Riess:2021jrx}, and being in agreement with $ H_0 = 69.96 \pm 1.05 \, \text{km}/\text{s}/\text{Mpc}$ from CCHP \cite{Freedman:2024eph}. This difference between these two local observations implies that there are some unknown factors to influence the local determination of $H_0$. In addtion, HZ-$\Lambda$CDM gives that $S_8=0.7645\pm0.0094$, which has no tension with the late universe observations \cite{KiDS:2020suj,Heymans:2020gsg}. But in the more recent KiDS-Legacy group’s results \cite{Stolzner:2025htz}, they found constraints consistent with Planck measurements with $S_8 = 0.814^{+0.011}_{-0.012} $, which implies that we might need to re-examine the $ S_8 $ tension.

Then we consider two simplest extensions of $\Lambda$CDM, adding the lensing amplitude $ A_\text{L} $ or the curvature density $ \Omega_\text{k} $ as a freedom parameter respectively. There are still CMB anomalies in these two extended models by using the HZ primordial power spectrum. The values of $ A_\text{L} $ and $ \Omega_\text{k} $ are in agreement with predictions from $\Lambda$CDM in the case of  the PL primordial power spectrum. Adding the \texttt{DESI BAO} data, the $ \Omega_\text{k} $ tends to be larger, even positive in the PL-$\Lambda$CDM model. Luckily, the level $ \lvert \Omega_\text{k} \rvert \sim \mathcal{O}(10^{-3}) $ is acceptable in the inflationary paradigm.

Finally, we show that the CMB angular power spectrum can be suppressed at large scales and uplifted at small scales by the HZ initial condition. The gravitational lensing and reionization can smooth the acoustic oscillations.  This effect can be mimicked by the combination of  \{$A_\text{s}, n_\text{s}$\}. Since the angular power spectrum is detected accurately at small scales, these uplifted effects from HZ spectrum should be offset by other approaches, like some smoothing effect from the gravitational lensing, reionization, or changing the density of baryon and CDM. Some of CMB anomalies or puzzles may be from degeneracies among these parameters \cite{Forconi:2023hsj,Forconi:2025zzu}. This implies that we need more independent measurements to constrain cosmological parameters, especially the gravitational lensing amplitude,  the optical depth of reionization, and the spatial curvature.

 Meanwhile, JWST and other local observations,  like the cosmic 21-cm signal, provide new methods to describe the EoR. In CMB analysis, the EoR is assumed by an instantaneous \textit{Tanh} model. This assumption prefers a smaller optical depth and thus may cause some cosmological tensions and CMB anomalies.  We will study the specific EoR from JWST measurements instead of a parameterized \textit{Tanh} model in the CMB analysis in the future.

\acknowledgments

We thank Gen Ye and Jun-Qian Jiang for helpful discussions. H.-H. Li wish to acknowledge the support of the China Postdoctoral Science Foundation, GZC20230902. T. Qiu is supported by the 36 National Key Research and Development Program of China with Grant No. 2021YFC2203100. J.-Q. Xia is supported by the China Manned Space Program with Grant No. CMS-CSST-2025-A01 and CMS-CSST-2025-A04.

\appendix
\section{A discussion of the goodness-of-fit}

The goodness-of-fit between PL-$\Lambda$CDM and HZ-$\Lambda$CDM models is shown in Table \ref{table:goofness-of-fit}. There is a large difference of $\chi^2$ between PL-$\Lambda$CDM and HZ-$\Lambda$CDM, as $\Delta \chi^2 = \chi^2_\text{min}(\text{HZ}) - \chi^2_\text{min}(\text{PL}) = 65 $. This huge positive value means that CMB observations disfavor the HZ spectrum as an initial condition. The HZ spectrum in extensions with the lensing amplitude $A_{\text{L}}$ or the spatial curvature density parameter $\Omega_{\text{k}}$ is also disfavored with $\Delta \chi^2 \sim 60$. The \texttt{Planck 2018 lowl.TT} and \texttt{ACT DR4} likelihoods show little favor to the HZ spectrum. The \texttt{Planck 2020 lollipop.lowlE} and \texttt{SPT3G} likelihoods show little favor to the PL spectrum. The most important source that causes this huge difference is the \texttt{Planck 2020 hillipop.TTTEEE} likelihood, the difference of $\chi^2$ is $60.1$. Then we choose the Planck 2018 likelihoods as a check. Choosing the CMB data sets as \texttt{Planck 2018 lowl.TT}(Commander), \texttt{Planck 2018 lowl.EE}(Simall), \texttt{Planck 2018 highl.TTTEEE}(Plik) and \texttt{SPT3G}, \texttt{ACT DR4} we used previously, there is still a large difference of $\chi^2$ between PL-$\Lambda$CDM and HZ-$\Lambda$CDM from the likelihood \texttt{Planck 2018 highl.TTTEEE}, as $\Delta \chi^2 = 61.2 $, which is similar to the result of \texttt{Planck 2020 hillipop.TTTEEE}.
\begin{table}[htpb]
\centering
\begin{tabular} {| l  | c | c | c | }
 \hline
 Parameter &  PL& HZ& $\Delta \chi^2$\\
\hline
$\chi^2_\mathrm{\Lambda\text{CDM}}                  $ & $32761.6\pm 9.7            $ & $32826.1\pm 9.6            $ & $ 65  $\\

$\chi^2_\mathrm{\Lambda\text{CDM},planck\ 2018\ lowl.TT\ clik}$ & $22.26\pm 0.73             $ & $19.802\pm 0.062           $ & $ -2.5  $\\

$\chi^2_\mathrm{\Lambda\text{CDM},planck\ 2020\ lollipop.lowlE}$ & $33.9\pm 1.4               $ & $37.1\pm 2.9               $ & $ 3.2  $\\

$\chi^2_\mathrm{\Lambda\text{CDM},planck\ 2020\ hillipop.TTTEEE}$ & $30519.7\pm 7.0            $ & $30579.8\pm 7.7            $ & $ 60.1  $\\

$\chi^2_\mathrm{\Lambda\text{CDM},SPT3G\ Y1.TTTEEE}$ & $1890.1\pm 6.2             $ & $1894.1\pm 6.5             $ & $ 4   $\\

$\chi^2_\mathrm{\Lambda\text{CDM},pyactlike.ACTPol\ lite\ DR4}$ & $295.7\pm 3.2              $ & $295.3\pm 3.0              $ & $ -0.4  $\\
\hline
$\chi^2_\mathrm{\Lambda\text{CDM}+ A_\text{L}}      $ & $32762.7\pm 9.6            $ & $32818.0\pm 9.8            $ & $ 55.3  $ \\
$\chi^2_\mathrm{\Lambda\text{CDM} + \Omega_\text{k}\text{(B)}}   $ & $32775.5\pm 9.7   $ & $32837.4\pm 9.8      $ & $ 61.9  $ \\
\hline
\end{tabular}
\caption{\label{table:goofness-of-fit} The goodness-of-fit between PL-$\Lambda$CDM and HZ-$\Lambda$CDM with total and individual likelihood. The last two rows are the extended cases. The difference $\Delta \chi^2 = \chi^2_\text{min}(\text{HZ}) - \chi^2_\text{min}(\text{PL})$. }
\end{table}

As has been pointed out in the study \cite{DiValentino:2018zjj}, HZ spectrum seems indeed strongly disfavored by Planck temperature and polarization data in the framework of $\Lambda$CDM. However, some extended scenarios like $ \Lambda $CDM$+ N_\text{eff} $ (the effective neutrino number) and $ \Lambda $CDM$+ Y_\text{He}$ (the helium abundance) with local observation data sets show no significant evidence against HZ spectrum. Therefore, it is still possible that HZ spectrum offers a possible alternative. The DESI DR2 data shown departure from $\Lambda$CDM model, preferring the dynamical dark energy and negative effective mass of neutrino \cite{DESI:2025zgx}.  A possible solution to increase the effective mass of neutrino is choosing a large optical depth \cite{Sailer:2025lxj,Jhaveri:2025neg,Cain:2025usc}. Our analyses show that the HZ spectrum tends to give a large optical depth. It is interesting to check whether the dynamical dark energy models or extended model with variation of neutrino mass support the HZ spectrum by adding the recent DESI DR2 BAO data.




\end{document}